\documentclass{article} 
\usepackage{colm2024_conference}
\usepackage{graphicx}
\usepackage{float}
\usepackage{url}

\usepackage{breakurl}
\usepackage[breaklinks]{hyperref}
\usepackage{xcolor}
\usepackage{tabularx}
\usepackage{array}
\usepackage{caption}
\usepackage{subcaption}
\usepackage{enumitem}
\usepackage{setspace}
\usepackage{comment}

\colmfinalcopy 
\begin{document}

\title{\centering \LARGE The infrastructure powering IBM's Gen AI model development}
\maketitle

\begin{center}
\textbf{Talia Gershon}$^\star$\quad
\textbf{Seetharami Seelam}$^\star$\quad
\textbf{Brian Belgodere}$^\star$\quad
\textbf{Milton Bonilla}$^\star$\quad
\textbf{Lan Hoang}\quad
\textbf{Danny Barnett}\quad
\textbf{I-Hsin Chung}\quad
\textbf{Apoorve Mohan}\quad
\textbf{Ming-Hung Chen}\quad
\textbf{Lixiang Luo}\quad
\textbf{Robert Walkup}\quad
\textbf{Constantinos Evangelinos}\quad
\textbf{Shweta Salaria}\quad
\textbf{Marc Dombrowa}\quad
\textbf{Yoonho Park}\quad
\textbf{Apo Kayi}\quad
\textbf{Liran Schour}\quad
\textbf{Alim Alim}\quad
\textbf{Ali Sydney}\quad
\textbf{Pavlos Maniotis}\quad
\textbf{Laurent Schares}\quad
\textbf{Bernard Metzler}\quad
\textbf{Bengi Karacali-Akyamac}\quad
\textbf{Sophia Wen}\quad
\textbf{Tatsuhiro Chiba}\quad
\textbf{Sunyanan Choochotkaew}\quad
\textbf{Takeshi Yoshimura}\quad
\textbf{Claudia Misale}\quad
\textbf{Tonia Elengikal}\quad
\textbf{Kevin O’Connor}\quad
\textbf{Zhuoran Liu}\quad
\textbf{Richard Molina}\quad
\textbf{Lars Schneidenbach}\quad
\textbf{James Caden}\quad
\textbf{Christopher Laibinis}\quad
\textbf{Carlos Fonseca}\quad
\textbf{Vasily Tarasov}\quad
\textbf{Swaminathan Sundararaman}\quad
\textbf{Frank Schmuck}\quad
\textbf{Scott Guthridge}\quad
\textbf{Jeremy Cohn}\quad
\textbf{Marc Eshel}\quad
\textbf{Paul Muench}\quad
\textbf{Runyu Liu}\quad
\textbf{William Pointer}\quad
\textbf{Drew Wyskida}\quad
\textbf{Bob Krull}\quad
\textbf{Ray Rose}\quad
\textbf{Brent Wolfe}\quad
\textbf{William Cornejo}\quad
\textbf{John Walter}\quad
\textbf{Colm Malone}\quad
\textbf{Clifford Perucci}\quad
\textbf{Frank Franco}\quad
\textbf{Nigel Hinds}\quad
\textbf{Bob Calio}\quad
\textbf{Pavel Druyan}\quad
\textbf{Robert Kilduff}\quad
\textbf{John Kienle}\quad
\textbf{Connor McStay}\quad
\textbf{Andrew Figueroa}\quad
\textbf{Matthew Connolly}\quad
\textbf{Edie Fost}\quad
\textbf{Gina Roma}\quad
\textbf{Jake Fonseca}\quad
\textbf{Ido Levy}\quad
\textbf{Michele Payne}\quad
\textbf{Ryan Schenkel}\quad
\textbf{Amir Malki}\quad
\textbf{Lion Schneider}\quad
\textbf{Aniruddha Narkhede}\quad
\textbf{Shekeba Moshref}\quad
\textbf{Alexandra Kisin}\quad
\textbf{Olga Dodin}\quad
\textbf{Bill Rippon}\quad
\textbf{Henry Wrieth}\quad
\textbf{John Ganci}\quad
\textbf{Johnny Colino}\quad
\textbf{Donna Habeger-Rose}\quad
\textbf{Rakesh Pandey}\quad
\textbf{Aditya Gidh}\quad
\textbf{Aditya Gaur}\quad
\textbf{Dennis Patterson}\quad
\textbf{Samsuddin Salmani}\quad
\textbf{Rambilas Varma}\quad
\textbf{Rumana Rumana}\quad
\textbf{Shubham Sharma}\quad
\textbf{Aditya Gaur}\quad
\textbf{Mayank Mishra}\quad
\textbf{Rameswar Panda}\quad
\textbf{Aditya Prasad}\quad
\textbf{Matt Stallone}\quad
\textbf{Gaoyuan Zhang}\quad
\textbf{Yikang Shen}\quad
\textbf{David Cox}\quad
\textbf{Ruchir Puri}\quad
\textbf{Dakshi Agrawal}\quad

IBM Research \\
\vspace{0.2cm}
\textbf{Drew Thorstensen}\quad
\textbf{Joel Belog}\quad
\textbf{Brent Tang}\quad
\textbf{Saurabh Kumar Gupta}\quad
\textbf{Amitabha Biswas}\quad
\textbf{Anup Maheshwari}\quad
\textbf{Eran Gampel}\quad
\textbf{Jason Van Patten}\quad
\textbf{Matthew Runion}\quad
\textbf{Sai Kaki}\quad
\textbf{Yigal Bogin}\quad
\textbf{Brian Reitz}\quad
\textbf{Steve Pritko}\quad
\textbf{Shahan Najam}\quad
\textbf{Surya Nambala}\quad
\textbf{Radhika Chirra}\quad
\textbf{Rick Welp}\quad
\textbf{Frank DiMitri}\quad
\textbf{Felipe Telles}\quad
\textbf{Amilcar Arvelo}\quad
\textbf{King Chu}\quad
\textbf{Ed Seminaro}\quad
\textbf{Andrew Schram}\quad
\textbf{Felix Eickhoff}\quad
\textbf{William Hanson}\quad
\textbf{Eric Mckeever}\quad
\textbf{Michael Light}\quad
\textbf{Dinakaran Joseph}\quad
\textbf{Piyush Chaudhary}\quad
\textbf{Piyush Shivam}\quad
\textbf{Puneet Chaudhary}\quad
\textbf{Wesley Jones}\quad
\textbf{Robert Guthrie}\quad
\textbf{Chris Bostic}\quad
\textbf{Rezaul Islam}\quad
\textbf{Steve Duersch}\quad
\textbf{Wayne Sawdon}\quad
\textbf{John Lewars}\quad
\textbf{Matthew Klos}\quad
\textbf{Michael Spriggs}\quad
\textbf{Bill McMillan}\quad
\textbf{George Gao}\quad\\

IBM Infrastructure \\
\vspace{0.2cm}
\textbf{Ashish Kamra}\quad
\textbf{Gaurav Singh}\quad
\textbf{Marc Curry}\quad
\textbf{Tushar Katarki}\quad
\textbf{Joe Talerico}\quad
\textbf{Zenghui Shi}\quad
\textbf{Sai Sindhur Malleni}\quad
\textbf{Erwan Gallen}\quad\\

Red Hat \\
\vspace{0.2cm}
$^\star$Corresponding Authors: \\
\texttt{tsgersho@us.ibm.com, sseelam@us.ibm.com, bmbelgod@us.ibm.com, bonillam@us.ibm.com}

\end{center}

\date{\today}

\begin{abstract}
AI Infrastructure plays a key role in the speed and cost-competitiveness of developing and deploying advanced AI models. The current demand for powerful AI infrastructure for model training is driven by the emergence of generative AI and foundational models, where on occasion thousands of GPUs must cooperate on a single training job for the model to be trained in a reasonable time. Delivering efficient and high-performing AI training requires an end-to-end solution that combines hardware, software and holistic telemetry to cater for multiple types of AI workloads. In this report, we describe IBM's hybrid cloud infrastructure that powers our generative AI model development. This infrastructure includes (1) Vela: an AI-optimized supercomputing capability directly integrated into the IBM Cloud, delivering scalable, dynamic, multi-tenant and geographically distributed infrastructure for large-scale model training and other AI workflow steps and (2) Blue Vela: a large-scale, purpose-built, on-premises hosting environment that is optimized to support our largest and most ambitious AI model training tasks. Vela provides IBM with the dual benefit of high performance for internal use along with the flexibility to adapt to an evolving commercial landscape. Blue Vela provides us with the benefits of rapid development of our largest and most ambitious models, as well as future-proofing against the evolving model landscape in the industry. Taken together, they provide IBM with the ability to rapidly innovate in the development of both AI models and commercial offerings.
\end{abstract}

\tableofcontents

\section{Introduction}
It is hard to overstate the important role of infrastructure in the successful development and efficient deployment of advanced AI models. Infrastructure selection and design impact the cost profile, speed, and efficiency of every stage of the AI life-cycle including data curation, pre-processing, tokenization, model training, adaptation, and inference. The emergence of generative AI and foundational models has led to a dramatic rise in the need for large-scale compute clusters with thousands of GPUs, which can be used together to train large models faster, provided that a sufficiently high-performing network and storage are available. Thus, the availability of a large-scale contiguous and high-performing infrastructure can have a significant impact on the time-to-value in the development of advanced models. In addition to hardware selections, the software we use to manage the infrastructure can also have a significant impact on time-to-value and the overall cost of achieving desired AI outcomes.


While it is known in the industry that state-of-the-art models are generally trained at scale over high-performance infrastructures (e.g. as described in papers from Meta ~\citep{touvron2023llama2openfoundation}, ~\citep{llama3}), few publications provide technical details on the design and operations of these systems. This document details IBM’s hybrid-cloud-based approach to building world-class infrastructure to support IBM’s model development activities at scale. This approach includes (1) The design and integration of AI-optimized supercomputing capabilities directly into IBM’s Cloud to deliver scalable, dynamic, multi-tenant and geographically distributed infrastructure for large-scale model training and other AI workflow steps and (2) the design and deployment of large-scale, purpose-built, on-premises hosting environments that are optimized to support our largest and most ambitious AI model training tasks. The former (Vela) provides IBM with the dual benefit of high performance for internal use along with flexibility to adapt to evolving commercial opportunities. The latter (Blue Vela) provides us with the benefits of rapid development of our largest and most ambitious models, as well as future-proofing against the evolving model landscape in the industry.

\section{Vela: An AI-optimized supercomputing infrastructure in IBM Cloud}

\begin{figure}[!htb]
   \captionsetup{justification=raggedright}  
   \begin{minipage}{0.95\textwidth}
   \centering
     \includegraphics[width=\linewidth,trim={-5cm 0cm 0cm 0cm},clip]{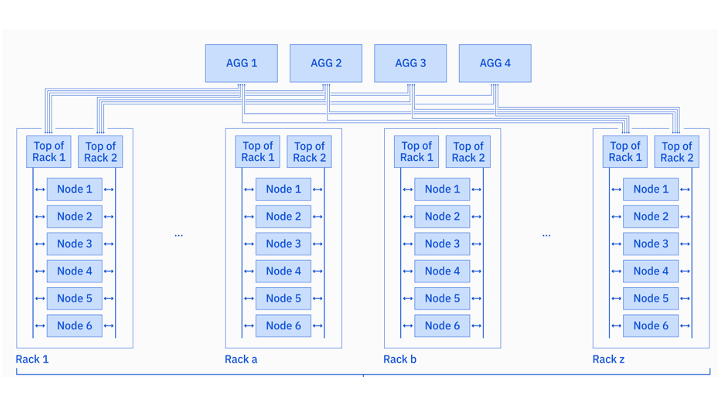}
     {(a) Overall system view}\label{Fig:rack}
   \end{minipage}\hfill
   \centering
   \begin{minipage}{0.95\textwidth}
   \centering
     \includegraphics[width=\linewidth,trim={0cm 0cm 0cm 0cm},clip]{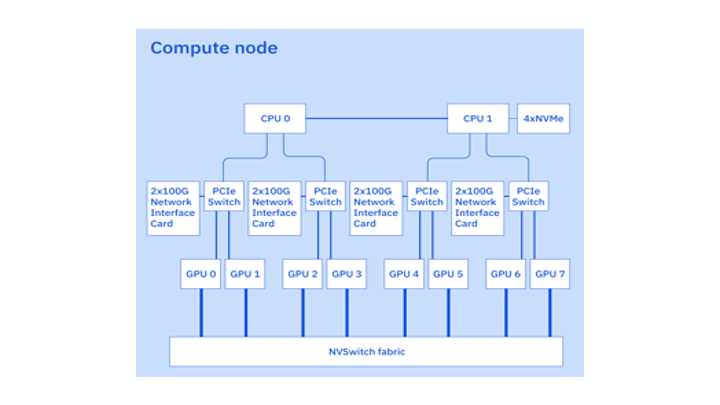}
    {(b) Compute node view}\label{Fig:node}
   \end{minipage}\hfill
    \caption{Vela System Architecture}\label{Fig:vela}
\end{figure}

In early 2023, IBM shared architectural details and design principles behind Vela, our first cloud-native AI-optimized supercomputer natively integrated into the fabric of IBM Cloud~\citep{vela-intro}. Some of these details are shown in Figure~\ref{Fig:vela}. Vela was designed to be flexible and scalable, capable of training today’s large-scale generative AI models, and adaptable to new needs that may arise in the future. It was also designed such that its infrastructure could be efficiently deployed and managed anywhere in the world. The following sections describe some of the technology and innovations that enable Vela's high performance, its flexibility, and its operational resilience.

\subsection{Vela Architecture}
Vela is a horizontally scalable data center system with two-layer spine-leaf CLOS architecture. 
Figure~\ref{Fig:vela}(a) shows the system/rack-level view of Vela and Figure~\ref{Fig:vela}(b) shows the architecture of the GPU nodes in Vela. Each of the node has eight 80GB A100 GPUs, which are connected to each other by NVLink and NVSwitch. Vela first came online in 2022 with GPU nodes containing 2nd Generation Intel Xeon Scalable CPU processors (Cascade Lake). A year later, it was expanded by a factor of roughly 2x where the new nodes contained 3rd Generation Intel Xeon Scalable processors (Ice Lake). Each node has 1.5TB of DRAM, and four 3.2TB NVMe drives. We anticipated that large memory and storage configurations would be important for caching AI training data, models, other related artifacts, and feeding the GPUs with data to keep them busy. To support distributed training, the compute nodes are connected via four 100G network interfaces that are connected in a two-level Clos structure as shown in Figure~\ref{Fig:vela}(a). 

To support high availability, which is  especially important for production cloud services such as watsonx.ai~\citep{ibm-watsonx}, we built network redundancy into the system. Each port of the network interface card (NIC) is connected to a different top-of-rack (TOR) switch (as shown in Figure~\ref{Fig:vela}(a)), and each TOR switch is connected via two 100G links to four spine switches providing 1.6TBps cross rack bandwidth and ensures that the system can continue to operate despite failures of any given NIC, TOR, or spine switch.

As shown in Figure~\ref{Fig:vela}(a), Vela racks have six servers, where the industry norm is between 2 and 4~\citep{meta-ai-clusters}. Our typical cloud racks are rated to provide 20kW of power to each rack from each of two redundant power distribution units (PDUs) (for a total of 40 kW available). Each GPU server draws a maximum of 6kW of power. Therefore, three servers can be accommodated per rack while preserving full power redundancy. The system, however, can accommodate greater density if a mechanism is in place to address potential failures of a power supply unit (PSU), where the servers are automatically throttled down to avoid overloading the healthy PDU. In order to enable this, we worked with our partners to develop a highly optimized power brake solution. When a PSU fails, the updated server firmware applies the power brake solution in about 2 seconds and the system throttles down to 3.2kW (taking each GPU from roughly 400W to 150W). Healthy PDU circuit breakers can tolerate power surges of up to 5 seconds. This allows us to essentially “overcommit” the amount of power available to each Vela rack safely. After extensive testing that showed all pertinent components were working safely with six servers per rack, without detrimental impact to the system or the workloads running on Vela, we proceeded to double the GPU density on Vela in 2023 within the same footprint originally allocated for the system in 2022. 


\subsubsection{Network}

Vela was designed for large model training. To support bigger models, trained over ever-larger data sets, moving faster means using more GPUs per job. As more GPUs compute in parallel, we need commensurate network performance to ensure that GPU-to-GPU communication doesn’t become a bottleneck to workload progress. 
We deliver this high performance networking by enabling two key technologies:  Remote Direct Memory Access (RDMA) over Converged Ethernet (RoCE), and GPU-direct RDMA (GDR). 

\begin{figure}[!htb]
 \centering
  \fbox{\includegraphics[width=0.9\linewidth,clip]{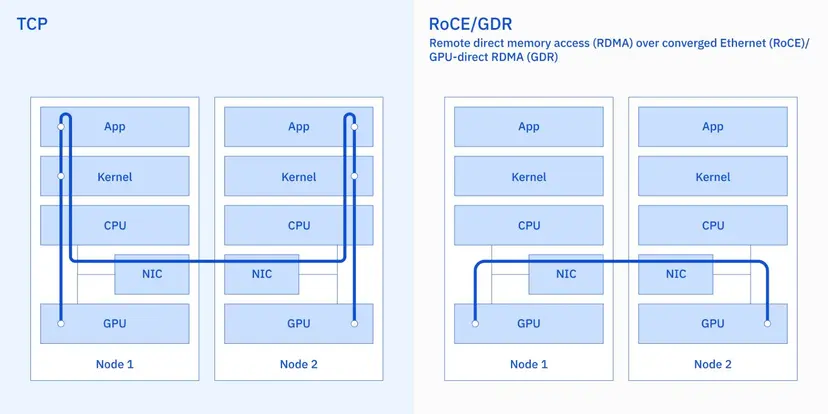}}
 \caption{Communication path with TCP and GPU Direct RDMA communication}
       \label{fig:tcpvsrdma}
 \end{figure}

RDMA allows one processor to access another processor’s memory without having to involve either computer’s operating systems. This leads to much faster communication between the processors by eliminating as many of the intervening processes as possible. GPU-direct RDMA (GDR) with ROCE, allows GPUs on one system to access the memory of GPUs in another system, using network cards (as shown in the Figure~\ref{fig:tcpvsrdma}), going over the Ethernet network.

We knew that AI training workloads would benefit from low latency RDMA communication paths on the nodes as shown in Figure~\ref{fig:tcpvsrdma}, resulting in more tokens processed per day and faster job completion times compared to TCP based communication. We also knew that achieving these benefits in a cloud native manner over a shared network infrastructure would come with its own challenges. While TCP communication works well over lossy, multi-pathed, Ethernet-based networks, network sensitive RDMA traffic requires quality of service (QoS) mechanisms to be supported by the underlying network infrastructure. We also knew we needed to offload as much network functionality as possible to the available hardware.

%

For RoCE to work in practice in our Ethernet-based production cloud, we needed to implement a robust congestion management mechanism.
Hence, we deployed a mechanism that relies on isolating RoCE traffic in a traffic class, monitoring congestion experienced by this class in the network, and notifying the senders to reduce their traffic to alleviate congestion before packet losses occur.
This process works by first marking the type of service (ToS) byte in the header of RoCE packets at the source. The first six bits of the ToS byte are reserved for the Differentiated Services Code Point (DSCP) tag and the last two bits are reserved for the Explicit Congestion Notification (ECN) tag.
We mark RoCE traffic with a specific DSCP tag and an ECN value to enable the switches to process RoCE traffic via a dedicated queue and to indicate the occurrence of congestion in this queue in the ECN field, respectively.
Second, Vela network switches are configured to monitor the buildup in the RoCE queue. 
They determine if there is congestion using a formula based on the length of the RoCE queue and mark the ECN field accordingly. 
Third, upon receiving a RoCE packet with congestion indication, the receiver sends a high priority message back to the sender known as the Congestion Notification Packet (CNP). Vela switches are configured to route such packets with high priority.
Finally, upon receiving a CNP packet, the sender throttles its traffic injection rate to reduce congestion.
The formula that the switches use to detect congestion of RoCE traffic is key to the effectiveness of this mechanism. In the Vela network, we tuned this formula to minimize packet losses.

In 2023, we enabled RoCE and GDR on Vela ~\citep{vela-upgrade}. This upgrade, which was several years of research in the making, required simultaneous changes and enhancements to nearly every part of our cloud stack, from the system firmware to the host operating system, to virtualization, to the network underlay and overlay. 

At the host level, new kernel drivers were needed with single-root I/O virtualization (SR-IOV) and RDMA support. At the network overlay level, IBM’s proprietary software-defined network (SDN) was extended to provide RoCE hardware offloads and marking of RoCE traffic with a specific DSCP label which would be recognized by the network switches for traffic isolation. We also needed to enable ECN in the SDN. Network interface cards (NICs) on the Vela nodes were configured to enable congestion control mechanisms. The network underlay on Vela was configured with a performance-tuned quality-of-service (QoS) policy, which ensured that RoCE traffic identified with the DSCP marking on the packet headers would be isolated in its traffic class and that an appropriate amount of bandwidth would be allotted to RoCE traffic during times of congestion. Vela network switches were configured to monitor the RoCE traffic class and mark the ECN bits of outgoing RoCE packets to indicate congestion when queue buildup occurs. The congestion control mechanism built into the Vela cluster relies on any NIC receiving ECN marked packets to notify the sender to throttle before packet losses occur.

The RoCE deployment in the Vela cluster was tuned to perform well with equal cost multiple-path (ECMP) routing.  The tuning effort included the QoS profile, the congestion control mechanism, switch buffers, and application properties such as flow characteristics. This end-to-end tuning was based on extensive studies conducted in a Research lab cluster built and configured like the Vela network.

In order to demonstrate the impact these changes have on network and AI workload performance, we captured data with communication microbenchmarks that represent the communication patterns of real workloads and also with a range of real workloads we expect our researchers to run on this system.

\begin{figure}[!htb]
   \begin{minipage}{0.85\textwidth}
     \centering
     \includegraphics[width=\linewidth,clip]{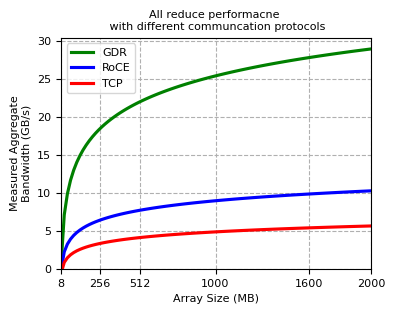}
     \caption{Performance of NCCL All Reduce collective with TCP, ROCE and GDR protocols}\label{fig:tcpvsrocedr}
   \end{minipage}\hfill
   \begin{minipage}{0.85\textwidth}
     \centering
     \includegraphics[width=\linewidth,clip]{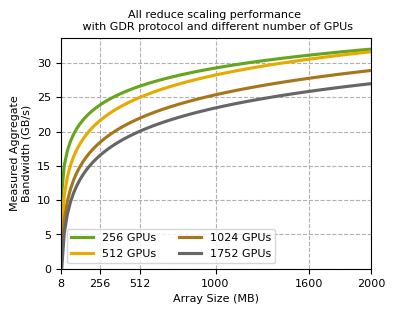}
     \caption{Performance of NCCL All Reduce collective with different number of GPUs}\label{fig:gdrscaling}
   \end{minipage}
\end{figure}

Figure~\ref{fig:tcpvsrocedr} shows the NVIDIA Collective Communication Library (NCCL) \texttt{all\_reduce} bandwidth performance as a function of the message size using TCP, ROCE, and GDR communication protocols over 1024 GPUs. For 8MB messages, GDR provides 2GB/s bandwidth, where TCP provides 0.2GB/s, a 10x difference. This improvement in bandwidth for smaller messages is primarily because of the latency improvements due to the shorter paths taken by the packets in the GDR case, as shown in in Figure~\ref{fig:tcpvsrdma} and the reduction on protocol overhead with direct memory copy. For 500MB and larger messages, GDR provides more than 20GB/s, and as high as 30GB/s bandwidth, whereas TCP saturates around 6GB/s, a 3-5x difference. With TCP, the link that connects the CPU and NIC becomes a bottleneck as it is traversed twice, once to copy the data from the GPU to the CPU and another time to copy the data from the CPU down to the NIC (see Figure~\ref{fig:tcpvsrdma}). With GDR, the data is directly passed from the GPU to the NIC eliminating the bottleneck link.

Figure~\ref{fig:gdrscaling} shows the scaling of network bandwidth performance as a  function of the number of GPUs (from 32 to 1752) participating in the collective operation using a ring algorithm. These results confirm that GDR results in scalable performance across a range of message sizes, from 8MB to 2GB, and a range of GPU counts, from 32 to 1752. These message sizes and the GPU counts are typical for AI training workloads in our system. Our AI researchers use these network scaling curves to guide how many GPUs to use for training and estimate the training times of jobs for different models they plan to train on Vela.

\subsubsection{Node Virtualization}
\label{node-virtualization}
When we designed Vela, we considered two consumption models: either make each node provisionable as bare metal (BM), or enable configuration of the node as a virtual machine (VM). It’s generally accepted that bare metal is the path to maximizing AI performance, but VMs provide more flexibility~\citep{vmvsbm}. Going the VM route would enable our service teams to easily provision and re-provision the infrastructure with different software stacks required by different stages of the AI workflow.  VMs would make it easy for our support team to flexibly scale AI clusters dynamically and shift resources between workloads of various kinds in a matter of minutes. For example, Vela nodes are constantly moved between training and inferencing clusters depending on the demand for resources. The downside of virtualization, historically, is that it reduces node performance~\citep{walkup2022best}. The research question to answer was: how close to bare metal performance could we achieve inside a VM? 

PCI-E device passthrough is the mechanism by which virtual machines directly access the physical devices installed on the host system. As a result, applications running inside VMs can take advantage of the full power of devices like GPUs and enable performant execution of applications such as gaming, CAD, and machine learning. We hypothesized that we could approach the bare metal performance of our GPU nodes by optimizing the way we pass the GPUs and virtual network functions into our VMs using Linux KVM.

\begin{figure}[!tb]
   \begin{minipage}{0.85\textwidth}
     \centering
     \includegraphics[width=\linewidth,clip]{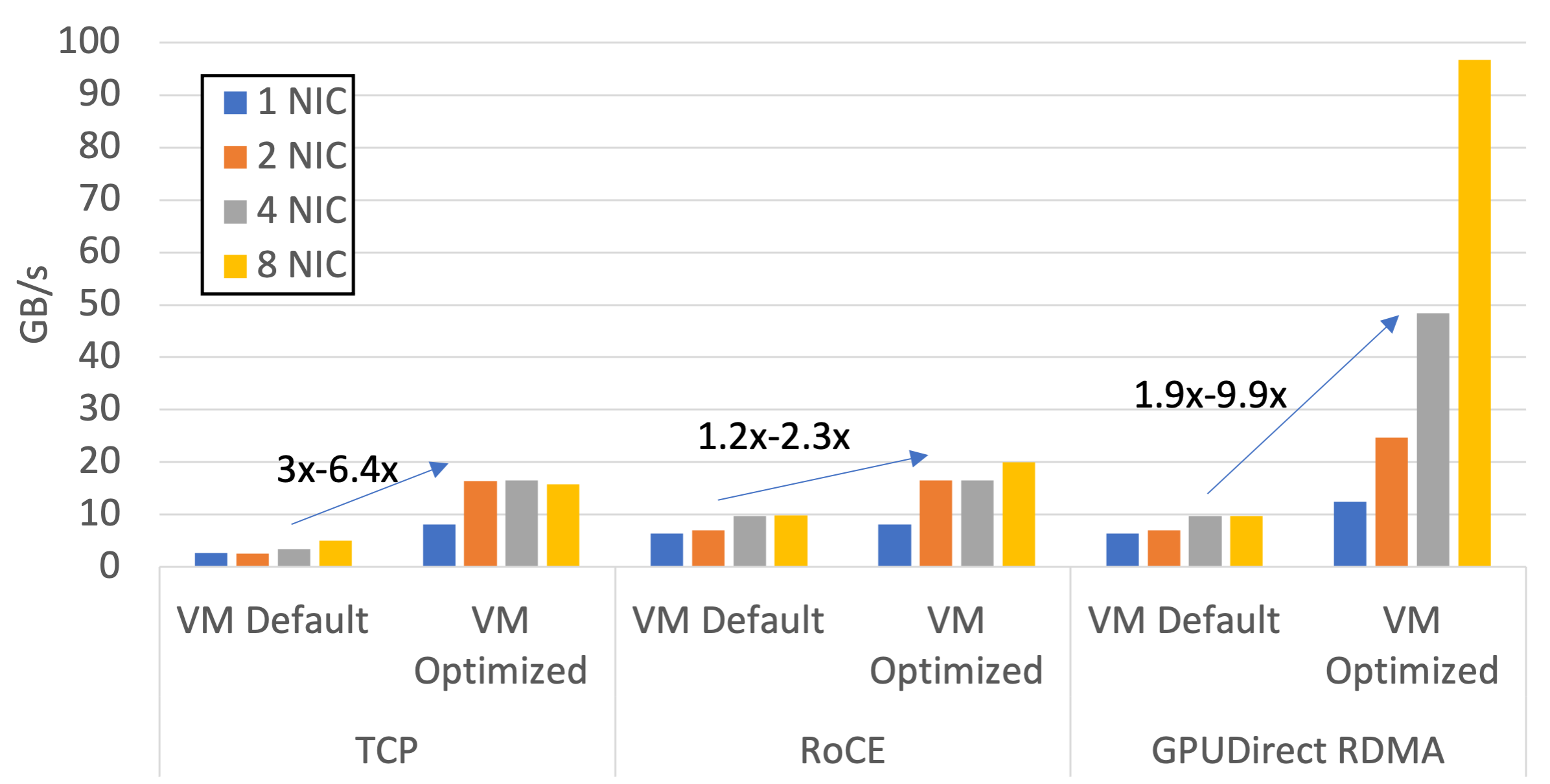}
     \caption{Performance improvement with optimizations}\label{fig:fig3}
   \end{minipage}\hfill
    \vspace{1cm}
   \begin{minipage}{0.85\textwidth}
     \centering
     \includegraphics[width=\linewidth,clip]{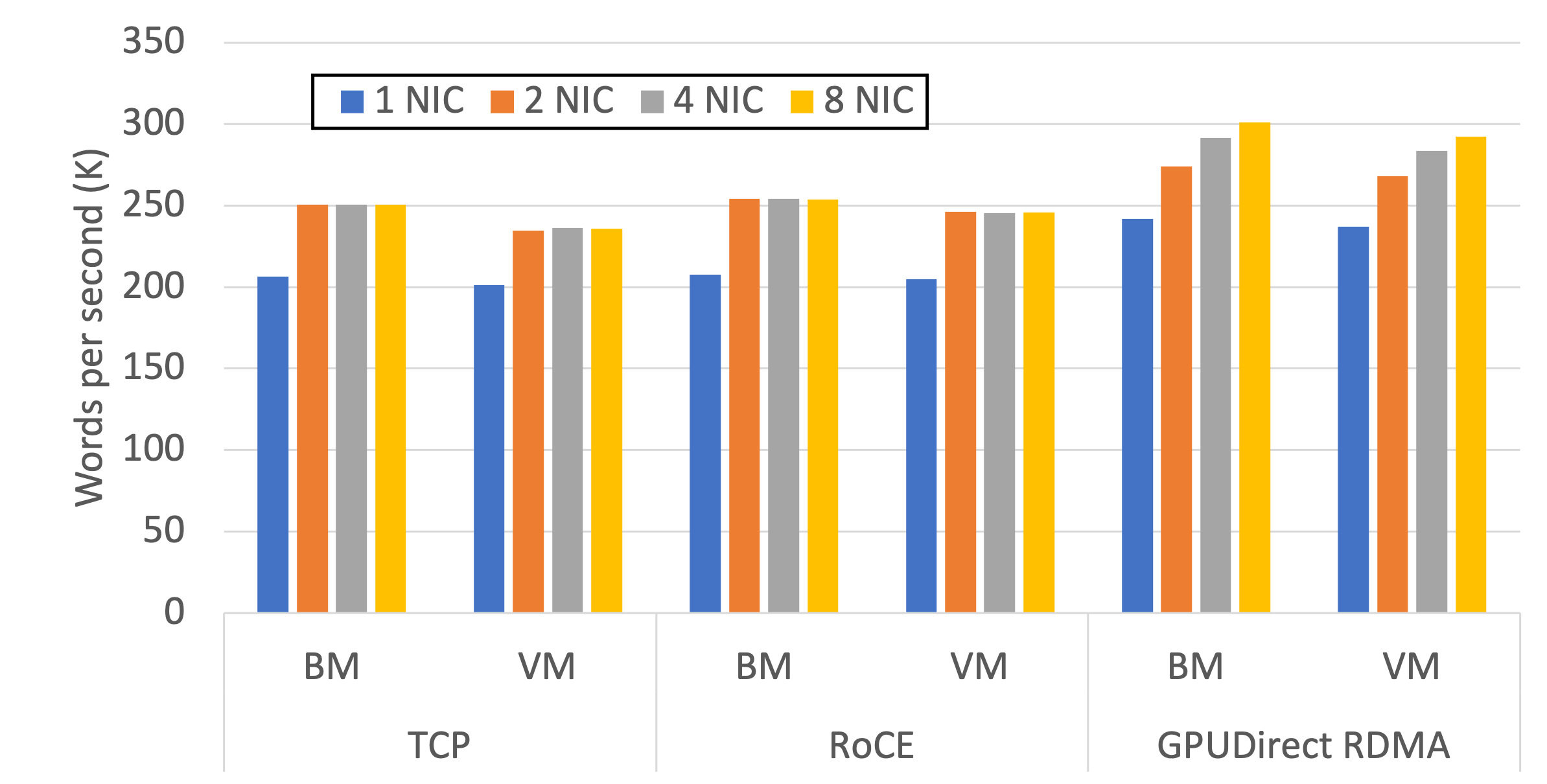}
     \caption{Virtualization overhead for NMT-12 two-node training job}\label{fig:f4}
   \end{minipage}
\end{figure}

As shown in Figure~\ref{Fig:vela}(b), modern AI compute nodes tend to have complex intra-node topology with multiple PCI-E switches connecting GPUs, network cards, and CPUs. These nodes are therefore challenging to virtualize with no performance loss. Before we began this optimization, we observed poor out-of-the-box AI workload performance and network performance (see Figure~\ref{fig:fig3} VM Default columns). We provisioned and compared KVM-QEMU-based VMs for TCP, RoCE, and GDR communication models with default configuration and optimized configuration. The optimized configuration achieves 2-10x improvement in network performance over the default configuration (Figure \ref{fig:fig3}).

Enabling a performant VM-based execution environment for AI workloads required configuration changes at different system layers~\citep{gtc22}. Specific optimizations were made in 
\begin{itemize}
    \item  the system BIOS (Virtual Machine Extensions like enable IOMMU, ACS, and SR-IOV support), 
    \item  the network adapter (disable relaxed PCI ordering, increase maximum accumulative outstanding read bytes, and enable selective repeat, direct access to ATS from the NIC to GPU buffers using PCI-E peer-to-peer transactions, and ATS), 
    \item the hypervisor (enable NVFs, huge pages, ACS on the PCI controllers, and ATS on the NVFs, and increase maximum PCI read request size to 4KB), 
    \item the guest XML (enable huge pages, NUMA domains, Device-NUMA mapping, host-physical-bit model for large memory VMs, and ATS on PCI controllers), and
    \item the guest operating system configurations (increase maximum PCI read request size to 4KB).  
\end{itemize}

We implemented these ideas on a system with 8 NVIDIA A100-80GPUs and measured performance of various microbenchmarks and workloads on bare metal and virtual machines. Our results shows that the optimized VM configuration resulted in close to bare metal performance across all the experiments. For example, we ran the NMT-12~\citep{ott2018scaling} model training job on single-node in VM and BM with the same configuration to demonstrate the virtualization overhead on real-world applications. The result shows that we can achieve 147K words-per-second (WPS) on BM and 140K WPS inside the VM, i.e., a virtualization overhead of about $~$5\%. We also evaluated BM and VM performance with Cuda Samples~\citep{cudasamples}, BERT-Large~\citep{devlin2019bert}, MegaTron~\citep{shoeybi2020megatronlm}, and T5 11B~\citep{raffel2020exploring}, and the overhead of VM ranges from "less than 0\%" to a maximum of 5\%. By "less than 0\%" we mean that VM execution can actually be faster; this is due to configurations such as large pages typically set for VMs, which is not the case in BM.

Virtualization also plays an important role in multi-node communication. To evaluate the virtualization optimizations for network, we used two compute nodes, each with  8 NVIDIA A100-80GB GPUs and 4 Mellanox ConnectX-6 Dx dual-port cards (i.e. 800 Gbps aggregate bandwidth), with the IBM Cloud KVM hypervisor (Linux 5.4 at the time) and guest running Ubuntu 20.04 (with Linux 5.4) operating system, and the latest software stack from NVIDIA and Mellanox. To benchmark the network performance, we executed \texttt{all\_reduce\_perf} from {\it nccl-tests} (a micro-benchmark suite provided by NVIDIA) to evaluate the network performance with TCP, RoCE, and NVIDIA GPUDirect RoCE protocols. Figure~\ref{fig:fig3} shows that the optimizations/changes we applied at different system layers improved the network performance significantly. We also executed the distributed NMT-12 model training job and demonstrated that we can achieve similar performance in VM and BM environments as shown in Figure~\ref{fig:f4}. We tested several other distributed AI training jobs, and the performance loss is less than $~$5\% in general.

In summary, we devised a way to expose all of the capabilities on the node (GPUs, CPUs, networking, and storage) into the VM and develop optimized VM configurations so that the virtualization overhead is less than 5\%, which is the lowest overhead in the industry that we’re aware of. These include configuring the bare-metal host for virtualization with support for Virtual Machine Extensions (VMX), SR-IOV, and huge pages. We also needed to faithfully represent all devices and their connectivity inside the VM, such as which network cards are connected to which CPUs and GPUs, how GPUs are connected to the CPU sockets, and how GPUs are connected to each other. These optimizations are part of our cloud control plane which creates the GPU VMs on the bare metal hosts. These, along with other hardware and software configurations, enabled our system to achieve close to bare metal performance.

\subsubsection{Storage}

During a large-scale model training job, several data- and I/O-intensive steps occur that can become a bottleneck to the overall training job progression in the absence of an optimized storage solution. The first one occurs when data, originally stored in an object store, must be accessed by the GPUs so that they can begin to compute. Loading data directly from object storage to each GPU's memory is slow due to the limited IOPs supported by typical cloud object storage. This bottleneck occurs both at the very beginning of the training job as well as every time the job stops and needs to be re-started again. As discussed later, component failures make this starting and stopping of jobs inevitable. A second I/O-intensive step of a model training job occurs during model "checkpointing". At periodic intervals during training, a record of the current set of model weights is sent to object storage, similar to occasionally "saving" the collective work of the GPUs. In both of these examples, a high-performance file system can be inserted between the object storage and the GPUs to act as an intermediating caching mechanism. In doing so, the data can be loaded into the GPUs much faster to start (or re-start) a training job, and model weights can be checkpointed to the file system at a much faster rate than when checkpointing directly to object storage. Thanks to unique technology in the file system we use, the checkpointed data can then be asynchronously sent to object storage but in a way that does not gate progress of the training job.

To realize these advantages, we use IBM Spectrum Scale~\citep{storage-scale}, which we will refer to as “Scale” in the following text. At the core of Scale is IBM's "General Parallel File System" (GPFS~\citep{storage-gpfs}), a highly successful and proven parallel file system with a strong high-performance computing heritage.

Scale is deployed in Vela using a disaggregated storage model. The dedicated Scale storage cluster consists of tens of IBM Cloud Virtual Server Instances (VSIs) with two 1TB virtual block volumes attached to each instance. The virtual block volumes are hosted on a next-generation cloud-native and highly performant block storage service in IBM Cloud that can meet the high throughput requirements of model training workloads. A single file system is created using all attached devices. Hence the total capacity of the file system accessible to Vela is hundreds of terabytes, which can be extended at any time as needed to petabytes.

As stated above, the majority of the data used by the training jobs running on Vela originates in IBM's Cloud Object Storage (COS). Similarly, some data produced on Vela, like model checkpoints, need to end up in COS for cost efficiency purposes. We configure Active File Management(AFM) technology~\citep{scale-afm} to transparently connect filesets to object storage buckets. File system namespaces represent objects in buckets as files and brings data from the object storage into the file system on demand. When a file is written to the file system, AFM eventually moves it to object storage. This means that our 140TB file system capacity is essentially acting as a read-write cache over potentially petabytes of cost-effective object storage. When the cache is full, AFM automatically evicts the data and metadata that was not recently used.

\begin{figure}[h!]
 \centering
  \fbox{\includegraphics[width=0.95\linewidth,trim={0.0cm 7cm 0 7cm},clip]{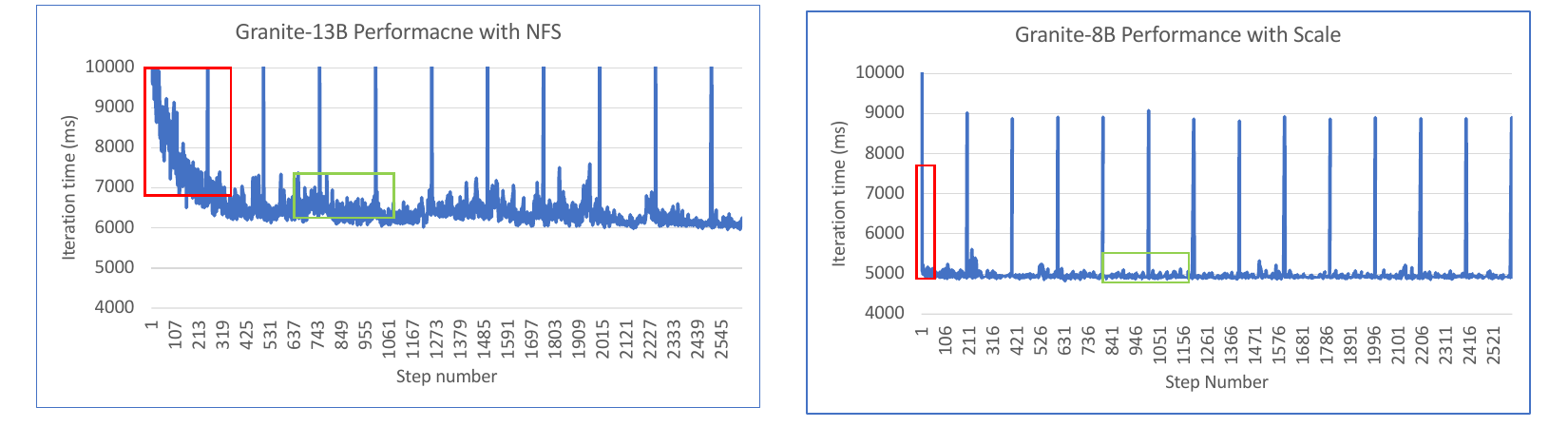}}
 \caption{Iteration time of an AI workload with NFS and Scale file systems. Iteration time improves by more than 10\% using the Scale file system.}
       \label{fig:spectrumscale-granite13b}
 \end{figure}

Before we deployed this storage solution based on Scale, AI researchers using Vela could either use IBM COS directly or an NFS file system that was deployed for Vela. Compared to NFS performance, our Scale file system achieves a near 40x read bandwidth speedup (1GB/s vs 40GB/s with Scale), which directly helps with input data read operations. Also compared to IBM COS bandwidth, the Scale file system achieves a near 3x write bandwidth speedup (5GB/s vs 15GB/s with Scale), which accelerates the checkpoint and other data write operations.  How do these numbers translate to real workload performance? Figure~\ref{fig:spectrumscale-granite13b} compares iteration times for an example Granite-13B AI training job using the NFS and another Granite-8B job using the  Scale file system. Several conclusions can be drawn from this comparison:
\begin{itemize}
    \item Because of the random I/O access patterns, concurrent accesses from multiple reads, and limited data reuse, the iteration time with NFS takes many steps to reach a steady state. In this experiment, it took more than 300 iterations. Because of the high bandwidth and low latency performance of Scale, and because of its ability to handle concurrent accesses, the iteration time reaches a steady state almost instantaneously. 
    \item During steady state training, multiple clients access the file system at the same time. Since NFS has limited concurrency support, the step time tends to vary by almost 50\% (e.g. from 6 seconds to 9 seconds). In the case of Scale, the step times varies between 4.8 seconds and 5.2 seconds, which is less than 10\% variation.
    \item Thanks to the overall higher performance of Scale, the step of the AI job is more than 10\% faster on average than using NFS, which directly reduces AI model training times by 10\%.
\end{itemize} 

\subsection{Vela Software Stack}

Vela is operated by IBM Cloud as IaaS (Infrastructure as a Service). On top of this, IBM Research and IBM Cloud manage different Red Hat OpenShift clusters which are used for tasks that span the entire AI lifecycle, from data preparation to model training, adaptation, and ultimately model serving. 

Leveraging the OpenShift platform for these use cases offers several advantages to AI researchers, AI service providers, and platform administrators. First, it allows AI researchers to bring their own container with software that is necessary to run their workloads. Some of our AI researchers use stable PyTorch versions, others use the Megatron framework, and others use nightly builds of the latest software. Bringing your own container allows AI researchers to define and run their own experiments on the platform without needing explicit coordination with system administrators. OpenShift also simplifies system-level monitoring and debugging using a rich set of out-of-the-box capabilities and a collection of specialized operators that will be described below, as well as automated job restart orchestrated by our job scheduler in the event of failures. 

For AI infrastructure service providers such as our cloud site reliability engineers (SREs), OpenShift offers a single platform to allocate appropriately-sized infrastructure to various services, for example hundreds of GPUs for training, individual GPU nodes for fine tuning, and a few GPUs for inference (which are further partitionable). This enables optimal use of resources for the workload and improves cost-efficiency of the infrastructure. 

For Platform administrators, OpenShift provides constructs (namespaces, quotas) to provision and manage resources among multiple users and projects with fine grained access controls. Platform level APIs allow the administrators to scale the system up and down based on resource availability and workload needs. For example, in Vela, resources are moved between OpenShift clusters for AI training and inference services based on business needs. 

While the OpenShift platform already comes with a rich collection of capabilities to expose high performance infrastructure to workloads (such as a GPU Operator, Network operator) and a sophisticated ecosystem of tools for system management (such as monitoring, logging, alerting), in the section below we will discuss some key capabilities we have developed based on the specific needs of workloads running on Vela and how we leverage them along with IBM Cloud capabilities to substantially enhance monitoring and diagnostic functions. 

\subsubsection{OpenShift Operators}

\textbf{Autopilot:}
IBM Research has created and open-sourced a tool called Autopilot~\citep{autopilot}, which is a cloud native observability tool implemented as a set of daemons, each one running on a GPU node. It implements a set of health checks that evaluate the status of the system. These health checks focus mainly on subtle/software issues, discussed more in the next section (i.e., row-remapping or PCI-E link degradation), but also run connectivity tests (i.e., ping, iperf) to verify that NICs are reachable. At the time of publication, the complete list of health checks includes the following: PCI-E bandwidth measurement between host and device, remapped rows evaluation, GPU power throttle enablement, all DCGM diagnostics, GPU memory bandwidth evaluation through DGEMM and DAXPY workloads, ping and iperf.
Any subset of health checks can be set up to run periodically and automatically on all nodes, but the user can also run them manually if needed. More extensive tests, namely DCGM diagnostics level 3, are also executed automatically only on nodes that have free GPUs. We added this deeper analysis because there have been episodes of subtle issues that are only revealed after running level 3 DCGM diagnostics, therefore we decided to run these more invasive health checks proactively and flag nodes with an ERR/PASS flag. Results from health checks are exported as Prometheus Gauges, so that users and admins can easily check the status of the system on Grafana.
This has significantly accelerated the discovery of issues in the cluster, and enabled proactive remediation.

\textbf{Multi-NIC CNI:}
As discussed in the previous sections, Vela GPU nodes have multiple 100G network interfaces and IBM Cloud uses single root I/O virtualization (SR-IOV)~\citep{lockwood2014sr} to expose multiple virtual interfaces per each physical interface. Multi-NIC CNI~\citep{multinic-cni} is a container-native interface built on top of Multus CNI with several important functions: 1. It discovers all of the interfaces on each host and handles them as a pool, 2. it assigns virtual interfaces for pods on top of the SR-IOV interfaces for TCP communication without encapsulation, and 3. it passes physical SR-IOV interfaces into the pods for GDR communication. These actions ensure that the workloads can achieve line rate network performance for TCP and GDR communication while code is running inside the pod. 

\textbf{CNSA:}
A Scale client cluster runs across Vela’s GPU nodes in container-native mode leveraging the CNSA edition of Scale~\citep{scale-cnsa}. It uses Kubernetes operators to deploy and manage Scale in a cloud-native fashion as well as a CSI Plugin for provisioning and attaching persistent volumes based on Scale. The client cluster does not contain any locally attached storage devices and instead performs remote mount of the file system in the storage cluster~\citep{scale-cnsa-remote-mount}. Such an architecture allows compute and storage clusters to grow and shrink independently as workload requirements change.

Data scientists gain access to the Scale file system using the traditional Kubernetes process. They create a Persistent Volume Claim (PVC) describing the size of the volume and the storage class referring to Scale. By invoking Scale’s REST API on the storage cluster, Scale’s CSI plugin creates a new fileset~\citep{scale-fileset-based-volumes}, which later can be attached as a volume to any pod running in OCP.

\subsubsection{Workload Performance on OpenShift}
\begin{figure}[h!]
    \begin{minipage}{0.85\textwidth}
    \centering
    \fbox{\includegraphics[width=\linewidth,trim={0.0cm 0cm 0 0cm},clip]{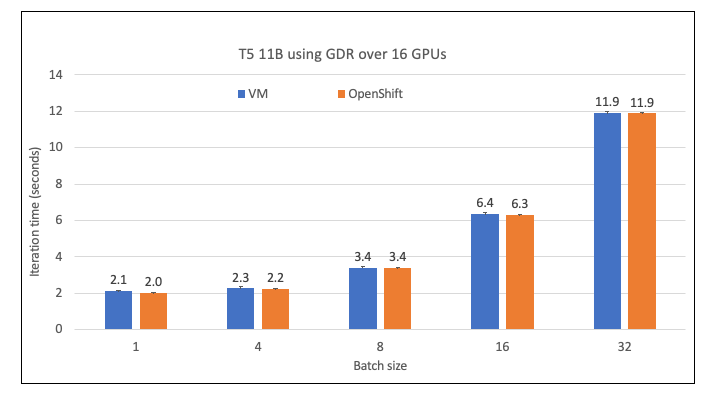}}
    \caption{Iteration time using virtual machines and OpenShift (lower is better). The performance difference is less than 4\%.}
       \label{vm-ocp-perf-comparison}
    \end{minipage}\hfill  
    \vspace{1cm}
    \begin{minipage}{0.85\textwidth}
    \centering
    \fbox{\includegraphics[width=\linewidth,trim={0.0cm 0cm 0 0cm},clip]{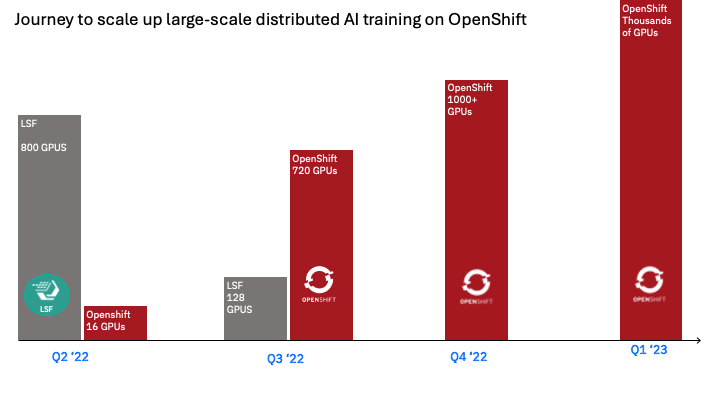}}
    \caption{Scaling OpenShift cluster from tens of GPUs to thousands of GPUs in nine months}
       \label{fig:ocp-scaling}
    \end{minipage} 
 \end{figure}

One of the concerns that is often expressed regarding the use of the OpenShift platform for performance-sensitive workloads is that it might introduce a resource overhead. We studied this by comparing the performance of various workloads running on OpenShift to those running directly in a virtual machines and found that the workload performance overhead is within the margin of error (i.e., under 5\%). Figure~\ref{vm-ocp-perf-comparison} shows a step time comparison for a representative AI workload running across 16 GPUs using VMs and OpenShift platform across a range batch sizes. The smaller batch sizes result in more communication per iteration, which allows the study of network overhead introduced by OpenShift. The larger batches result in more computation per iteration (hence larger iteration times), which allows us to study the container virtualization overhead. Across all of the batch sizes, the iteration times with OpenShift are within 4\% of the iteration time with VMs. Note that the step times are the same or even better with OpenShift compared to VMs in some cases. In the prior section on node virtualization, we summarized the overall virtualization overhead as varying between nearly 0\% to up to 5\%; the addition of an OpenShift layer does not meaningfully change this. The overall overhead is still contained within roughly 5\% relative to bare metal. 

OpenShift does in fact run more processes on the node compared to a typical HPC scheduler such as monitoring agents, network operators, logging agents, etc but their cumulative CPU usage is within 2\% and cumulative memory usage is under 4\%. This a reasonably small overhead and it can be ignored for all practical purposes.

When Vela first came online in 2022, we initially deployed with a traditional HPC scheduler called IBM Spectrum LSF and only a small (2 node) OpenShift cluster, as shown in Figure~\ref{fig:ocp-scaling}. We started to add capabilities like high performance networking using Multi-NIC CNI, optimized scheduling using the MCAD scheduler~\citep{ibm-mcad}, container-native storage using IBM Storage Scale CNSA, advanced monitoring and automated health checks using Autopilot, and grew the platform over nine months to support thousands of GPUs under a single cluster. To the best of our knowledge, this is the largest OpenShift cluster with GPUs in production anywhere in the world.

While a single cluster is good for resource consolidation, training workloads and inference workloads have distinct security and scaling requirements so we also deploy additional clusters on Vela for production watsonx.ai inference services. 

\subsection{Operational efficiency and resilience}

It is easy to understand the direct relationship between having a high-performance infrastructure and training a model as quickly as possible. Equally important is the need for technology to enable operational efficiency. This includes supporting dynamicity and variability in the types of experiments that researchers want to run as well as the need for technology to detect and address a whole host of inevitable failures that are guaranteed to occur when training large models on a large and complex infrastructure. This section details an operational and technological approach for rapid experimentation and rapid failure resolution, which are both equally critical for AI model training agility.

\subsubsection{Addressing component failures}
In distributed training, when hundreds or thousands of GPUs are used together to train a model, a reduction in the performance of even a single node can reduce the entire job's performance. This problem will persist until the slow node is identified, removed from the job, or fixed, and the job is restarted on fresh nodes with identically high performance.

Over the course of operating and training models on Vela over the last two years, we encountered a variety of issues that resulted in job slowdowns of anywhere from a few percent to over 3x. In some cases, failures caused jobs to crash altogether. We classify these failures into three buckets and describe the root causes and the mitigations we developed to handle the failures in Table~\ref{tab:failures}: 
\begin{itemize}
    \item Clear hardware failures,
    \item Subtle hardware failures, and
    \item Software failures.
\end{itemize}

\begin{table}
    \centering
     \setlength{\leftmargini}{0.4cm}
    \begin{tabular}{| p{0.28\textwidth} | p{0.31\textwidth} | p{0.31\textwidth} | }
        \hline
        Failure type & Root cause & Mitigations developed  \\
        \hline
         Hardware failures (host crash) & 
        \begin{itemize} 
            \item GPU HGX board failures
            \item Memory DIMM failure
            \item NVLink/switch failure 
        \end{itemize} & 
        \begin{itemize} 
            \item Slack alert on host crash
            \item Automatic VM restart
            \item Automatic job restart
        \end{itemize}  \\
        \hline
            Subtle hardware failures (no host crash) & 
        \begin{itemize} 
            \item Failure of GPUs
            \item GPU HBM Memory corruption 
            \item PCI-E link failure
            \item Port failures
            \item Power feed failure
        \end{itemize} & 
        \begin{itemize} 
            \item Slack alert on port, GPU, other critical component failures
            \item Alert based on host BMC logs
            \item Enhanced metrics collection via Autopilot
        \end{itemize}  \\
        \hline    
            Software failures  & 
        \begin{itemize} 
            \item PCI-E Link degradation
            \item Cuda memory allocation error
            \item HBM Memory row remaps
        \end{itemize} & 
        \begin{itemize} 
            \item Checks of PCI-E links
            \item Alerts based on application logs
            \item Periodic VM reboots
        \end{itemize}  \\
        \hline         
    \end{tabular}
    \caption{Infrastructure failure types, root causes, and mitigations.}
\label{tab:failures}
\end{table}

\textbf{Clear hardware failures:}
We classify a failure as a clear hardware failure when the host crashes. We observed that on average about $2\%$ of our hosts crashed per month over the past 2 years and as many as $5\%$ in the worst case. A primary source for these failures is the failure of the board that holds the GPUs (called an HGX baseboard). A second source, albeit less frequently observed, is the failure of the NVLink or the NVSwitch system. In both of these failure cases, the GPU system has to be removed from the cluster, and the system has to be fixed or replaced, in most cases by the vendor, before it can be returned back to the cluster. Another source of clear hardware failures have been the memory DIMM's of the host. These DIMM's can be replaced by our operations teams so that the system can return to operations relatively quickly compared to the other failures. 

When these kinds of failures occur, the host crashing causes the application to crash as well, which would normally require manual intervention to restart the application on a healthy set of nodes. To alert the system reliability engineers (SRE's), we developed Slack automation to send a message when the cloud control plane detects such as a host crash. In addition, if there are free hosts, our platform control plane (RedHat OpenShift host controller, which we developed) automatically restarts the VM so that the pool of resources provisioned in the cluster stays constant. Our job management system also automatically restarts the job on a healthy set of nodes, typically from a previous checkpoint. 

To make sure that we can always restart the training job with the same number of GPUs after every node failure, we keep approximately 10\% of the nodes as a buffer pool, where the buffer is replenished by adding new nodes or fixing issues on the existing nodes.

\textbf{Subtle hardware failures:}
These failures usually don't result in a host crash but they could result in application failure or slowdown in application performance. The root causes for these failures include the failure of one or more GPUs, corruption in GPU HBM memory cells resulting in incorrect outputs, or PCI-E link, network port, or power feed failures which impact the training speed of the application. 

For example, in one instance, a Granite-20B~\citep{granite_open} training job was running on 768 GPUs when suddenly the team noticed a performance degradation of almost 3x in step time. Debugging of all the 96 nodes of the job showed that one node had a power card issue, which resulted in that node automatically throttling down its power consumption to protect the GPUs and the healthy PSU. Recall that our custom power throttling reduces the GPU power from 400W to roughly 150W. This resulted in that node processing its computations roughly 3x slower than others, dragging down the performance of the entire job. In another instance, a network port on one of the jobs failed, which caused that node to send and receive data at a slower rate than others. Because each of our NICS has two ports and because the data is routed via ECMP across these ports, the loss of a single port didn't result in crash of the job but rather a slowdown. Once the problem node was removed from the job, the application performance was restored back to the original throughput. 

For each of these cases, we added additional alerts at the infrastructure layer (e.g. alerts based on messages from the host VM logs, application logs, and host BMC logs) so that AI researchers and system administrators can take proactive steps to rapidly locate the problem nodes and rapidly restart the job using an available pool of buffer nodes. 

\textbf{Software failures:}
The observation of various application failures or performance degradation has also been traced back to problems in the system software, such as the firmware on the GPU boards, the firmware on the network cards and PCI-E switches and user application code. As an example, the GPU nodes in Vela have PCI-E Gen3 and Gen4 links but on occasion performance resembling a lower PCI-E generation has been observed (e.g. ~4 GB/s, resembling PCI-E Gen 1). Such a degradation could impact application performance anywhere from a few percent to 2x. These PCI-E link downgrades are the most frequently observed failure among all the issues listed in this section. Fortunately this issue is resolved most of the time ($\geq95$\%) by resetting the PCI-E device, which is often done by rebooting the VM. Root cause for these degradations is still under investigation. 

Another software issue we have observed is associated with the "remapping" of HBM memory rows ~\citep{hbm-row-remap}, where row remapping is a hardware mechanism supported for NVIDIA A100 GPUs that replaces degraded memory rows with spare rows in the HBM bank. While this does not always immediately affect GPU workloads, a GPU reset is highly recommended as soon as possible by rebooting the virtual machine.

\subsubsection{System and workload monitoring}

In Table~\ref{tab:failures}, we outlined a variety of failures that could lead to costly and disruptive incidents across the entire infrastructure. This section discusses how we mitigate their impact. Specific types of failures, such as host crashes or GPU failures, pose the most significant threats to ongoing AI workloads. Typically, the workload stalls for a period before crashing, resulting in wasted GPU hours during the stall, after which the job is automatically rescheduled. However, simply restarting the job isn't always effective. Some GPU failures, undetectable by the OpenShift control plane, may leave defective nodes in the scheduling pool, potentially leading to their reuse upon job relaunch. In such cases, the quickest recourse is to analyze application logs to identify the failure. Given the scale of our cluster, comprising thousands of GPUs, where jobs utilize many hundreds of GPUs with thousands of processes/logs, these debugging processes consume considerable time. Hence, to mitigate these issues, we require monitoring capabilities that promptly alert stakeholders and pinpoint failing GPUs or nodes. This enables immediate actions to remove defective nodes and reliably restart jobs. Our approach combines IBM Cloud observability and monitoring services, including Activity Tracker, LogDNA and Alert Manager from OpenShift. This alerting mechanism obviates the need for time-consuming troubleshooting efforts.

\begin{figure}[h!]
 \centering
  \fbox{\includegraphics[width=0.9\linewidth,trim={0.0cm 3cm 1cm 3cm},clip]{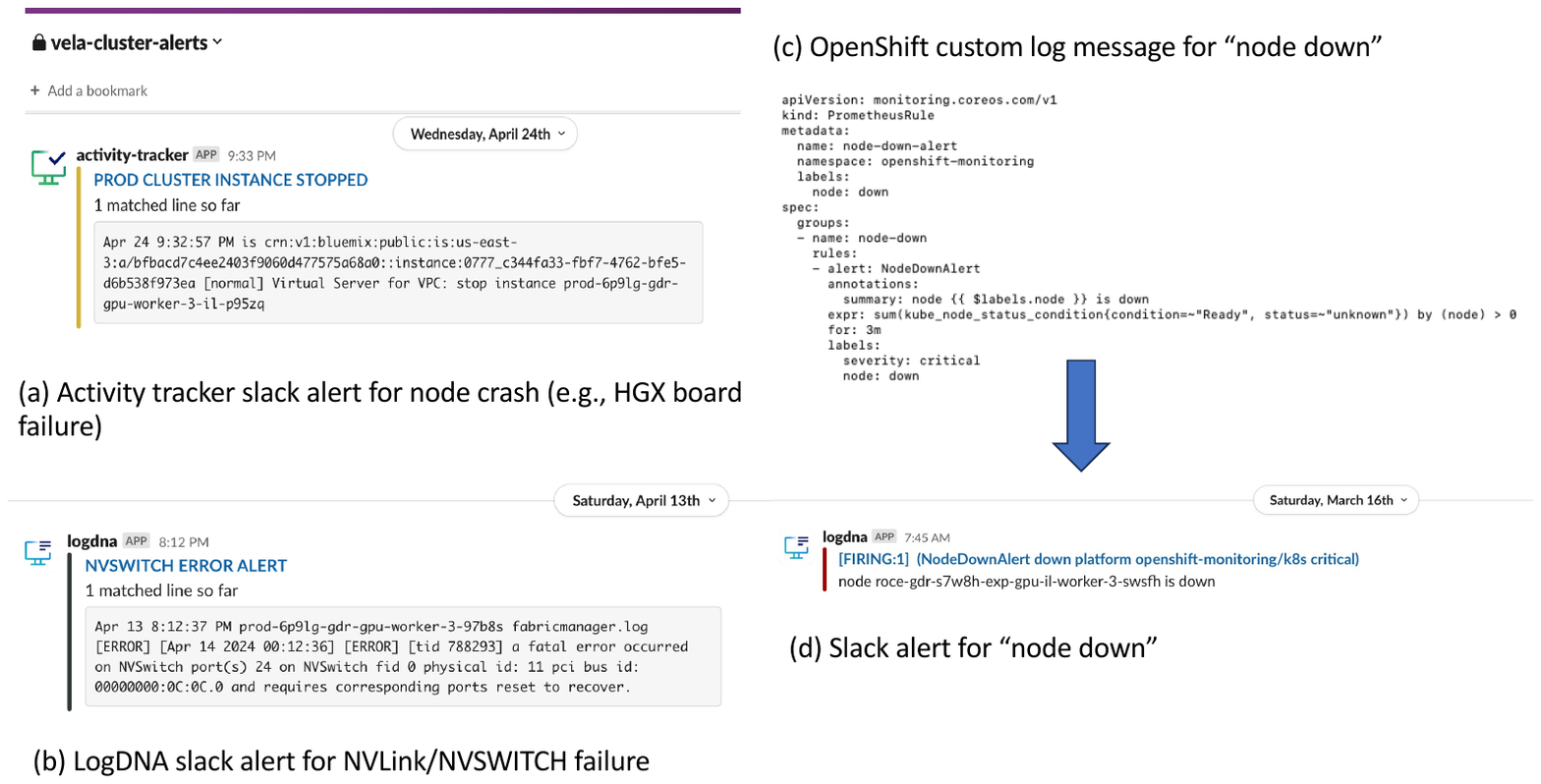}}
 \caption{Alerts for hardware failures and other node maintenance events. The alert are generated by IBM Cloud Activity tracker and LogDNA based on log messages from the hosts and custom messages generated by OpenShift monitoring rules.}
       \label{fig:hardfailures}
 \end{figure}

To detect hardware failures described in Table~\ref{tab:failures}, we utilize the IBM Cloud Activity Tracker service, which monitors the status of all virtual machines in the cluster. In the event of a node crashing and transitioning to a stopped status, Activity Tracker allows Slack integration so a Slack notification is automatically dispatched as shown in Figure~\ref{fig:hardfailures}(a). Utilizing the OpenShift Prometheus monitoring stack, we can accomplish a similar task as an alternate approach to using IBM Cloud Activity Tracker in cases where IBM Cloud is not available such as private cloud deployments. Figure~\ref{fig:hardfailures}(c) shows an example of a Prometheus rule to detect ``node down" and generate a custom log message, which is then received by Alertmanager in OCP. A Slack notification (shown in Figure~\ref{fig:hardfailures}(d) is sent whenever the rule is triggered. Since NVLink or NVSwitch system failures frequently result in host crashes as well, we also monitor Fabric Manager logs using LogDNA.  The logs are parsed and a Slack alert is generated (shown in Figure~\ref{fig:hardfailures}(b)) when  \texttt{NVSWITCH} and \texttt{fatal error} appear in the log. 

\begin{figure}[h!]
 \centering
  \fbox{\includegraphics[width=0.9\linewidth,trim={0.0cm 3cm 1cm 3cm},clip]{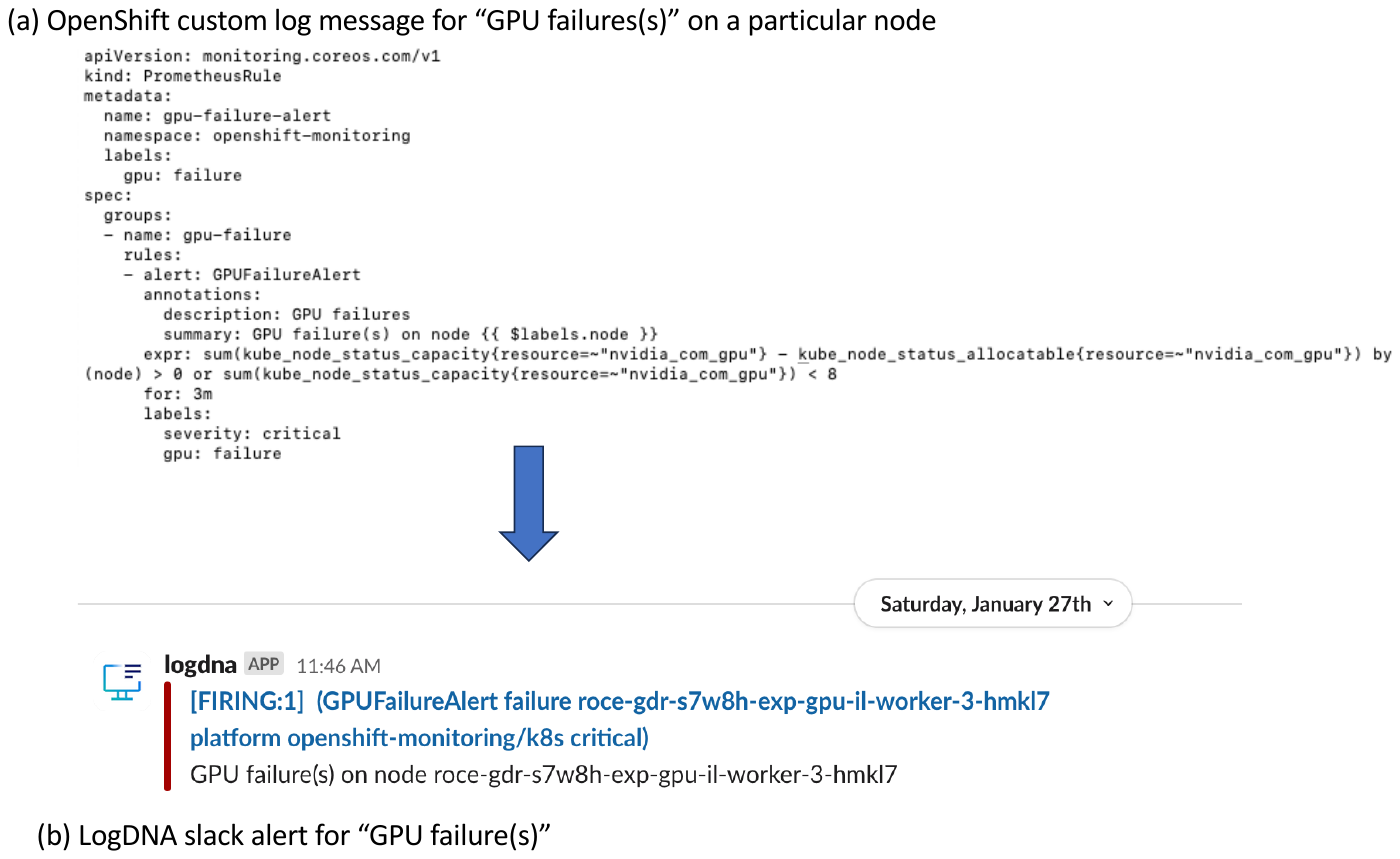}}
 \caption{An example custom alert for GPU failure event captured at the OpenShift layer. OpenShift monitoring generates a custom alert message, that triggers an event at LogDNA and an alert is generated to slack. In this case, the node is healthy but one or more GPUs have an issue that requires attention.}
       \label{fig:subtlefailures}
 \end{figure}

As mentioned earlier, not all GPU failures are detectable at the cloud AI infrastructure layer or within the OpenShift platform. In such cases, application logs offer valuable insights when encountering specific GPU failures. Rather than manually parsing through over 1000+ log files, we can once again rely on LogDNA to monitor application logs by matching CUDA error keywords. Slack alerts then precisely pinpoint which pod within the job is affected by the GPU failure. For example, to detect subtle hardware failures described in Table~\ref{tab:failures}, we apply a similar approach as above where we define custom monitoring rules at the OpenShift layer (as shown in Figure~\ref{fig:subtlefailures}(a)), process that event with LogDNA and generate an alert in Slack (as shown in Figure~\ref{fig:subtlefailures}(b)). 

To identify subtle hardware and software failures that could significantly impact end-to-end performance for AI workloads while jobs continue to run, we've adopted a proactive approach to streamline the root-cause analysis process. This approach starts with a set of microbenchmarks~\citep{autopilot} and tools to characterize various system components such as PCI-E links, NICs, GPUs, etc. We've categorized these tests into two groups: lightweight tests that can run concurrently with AI workloads, and more intrusive tests that require dedicated resources and are conducted when nodes are not in use by customers.
We aim to run lightweight tests periodically on every node, regardless of whether user jobs are present. In contrast, we run more intrusive tests only when nodes are not in use by customers. To facilitate this approach, we use the Autopilot and Multi-NIC CNI tools described above. The measurements from these tests are exported to Prometheus, allowing us to compose queries using Prometheus metrics to proactively detect any link degradation or node issues that may lead to gradual performance regression for AI workloads.

\begin{figure}[h!]
 \centering
  \fbox{\includegraphics[width=0.9\linewidth,trim={0.0cm 3cm 1cm 3cm},clip]{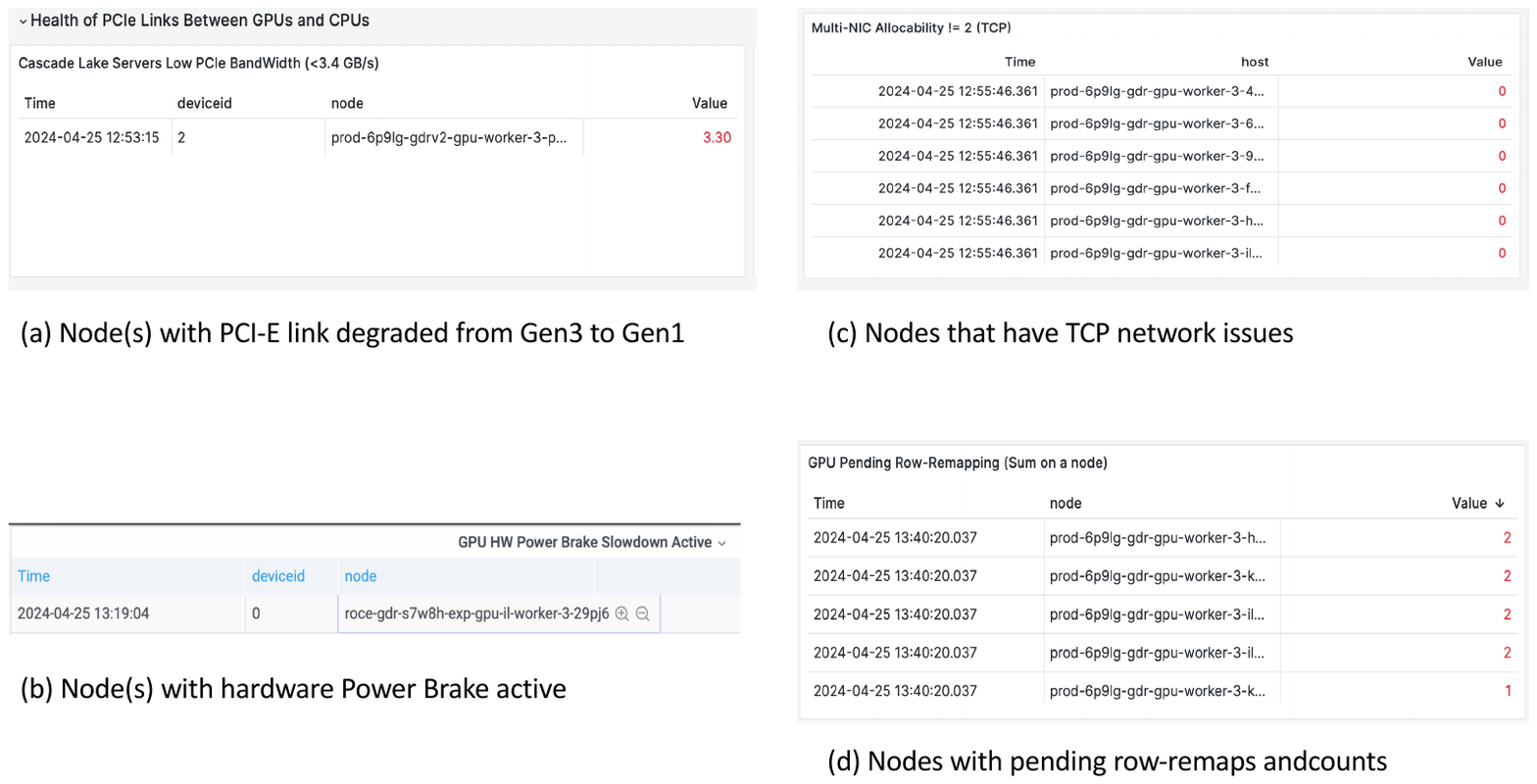}}
 \caption{Multiple dashboards showing (a) nodes with PCI-E link degradation, (b) nodes where power break is active, (c) slack alert for CUDA memory error, (d) nodes with TCP network connectivity issues, and (e) nodes with pending row-remap requests.}
       \label{fig:subtleandsoftwarefailures}
 \end{figure}

Given that some issues require deeper investigation, we rely on Grafana dashboards to flag potential nodes requiring detailed analysis. A subset of these dashboards are shown in Figure~\ref{fig:subtleandsoftwarefailures}. PCI-E link degradation (Figure~\ref{fig:subtleandsoftwarefailures}(a)) stands out as the most common issue leading to performance degradation. To address this, we utilize Nvidia's CUDA code and execute periodic tests through Autopilot. These tests are lightweight and minimally interfere with customers' workloads. To prevent false positives caused by both workloads and benchmarks contending for the same resources, we monitor values over an interval of 12 hours. Essentially, we sample 12 data points, collected every hour, and trigger alerts only when the average value falls below a certain threshold. This approach has proven highly accurate, effectively eliminating false positives and accurately identifying nodes experiencing link degradation. An example of a node where PIC-E link degraded below 3.4GB/s is shown in Figure~\ref{fig:subtleandsoftwarefailures}(a). These thresholds are specific to PCI-E link generations and users can define them as they see appropriate for their environment. 

Additionally, we've encountered several incidents stemming from power supply issues affecting the servers. Such issues result in GPU throttling, substantially reducing GPU compute power. Fortunately, by examining the GPU counters (such as power break slowdown) provided by the \texttt{nvidia-smi} tool we can efficiently identify nodes experiencing this problem without needing to conduct intensive tests on the system. This test is also integrated into Autopilot, and we can flag it in the Grafana dashboard, which shows up as the entry in Figure~\ref{fig:subtleandsoftwarefailures}(b).

Another frequently occurring issue is due to correctable errors in HBM memory of the GPUs. This shows up as GPUs currently undergoing pending row remapping in system messages. We leverage \texttt{DCGM\_FI\_DEV\_ROW\_REMAP\_PENDING} from the DCGM exporter in the Nvidia GPU operator. While customers may still utilize these nodes, it's strongly adviseable to reset these GPUs promptly. If applications push GPU memory usage closer to full capacity, the likelihood of job crashes increases significantly. Therefore, we've also established a panel on the dashboard (shown in Figure~\ref{fig:subtleandsoftwarefailures}(c)) to notify system administrators that these nodes are workload-free and can be reset. It's important to highlight that corrupted GPU memory failures can lead to silent errors at the application level. Applications may continue to run without apparent issues until the logs reveal inflated loss values during the training loop. These failures can occur at any point during training, resulting in a significant waste of GPU hours if not monitored for loss curve convergence. The DCGM diagnostics at levels 1\&2) are unable to detect this issue and level 3 diagnostics are needed, which requires exclusive access to the GPUs. To address this, Autopilot incorporates this test as part of the intrusive tests that are run when GPUs are not actively being used for AI workloads. The results are exported to both Prometheus and node labels for monitoring and analysis.

Advanced networking plays a crucial role in efficiently scaling up AI training with hundreds of GPUs. As mentioned earlier, our Multi-NIC CNI operator enables passing multiple secondary network interfaces to each pod, allowing us to leverage high-performance infrastructure capabilities without incurring performance overhead. The Multi-NIC CNI checks the health of TCP interfaces but RoCE/GDR interfaces are harder to monitor because once a job starts they are typically dedicated to client workloads. We have developed a new way to ensure that these interfaces are healthy. Our enhanced Multi-NIC health checker gathers node network bandwidth data for all 2-node pairs on every port, and appends this information to the \texttt{status} section of a custom resource object. This status section can be then be queried to detect nodes with degraded RoCE/GDR performance.

\textbf{Summary:}
Taken together, these capabilities have allowed our SRE's to go from a reactive posture (i.e. debug an issue whenever users report them) to a proactive posture (i.e. inform users when their jobs are experiencing issues related to platform or infrastructure even before users notice them) and dramatically reduce the time to root-cause the issues described in Table~\ref{tab:failures} by over 2x.

\subsubsection{Checkpointing}
Due to the unavoidable nature of various component failures in the system, some amount of time is always lost during model training. To limit the amount of work lost in the event of a failure, we configure our training runs to periodically checkpoint the model state to persistent storage. Since Vela is built into IBM Cloud, we used IBM COS to checkpoint the state every few hours. We use the famous Young's formula \citep{checkpoint-times} to compute the checkpoint interval, which is computed as 
$t_{checkpoint} = \sqrt{2\delta M}$ where 
$\delta$ is the time to checkpoint the model state, and
$M$ is the time between failures.

From the time that training starts, we measure that less than 10\% of the total time is lost due to failures, which includes time to checkpoint the data, time to recompute steps from a previous checkpoint in case of a failure, time to debug system issues and isolate faulty nodes. 
\\\\

\subsection{Workload performance on Vela} \label{sec:vela-workload}

Over the last two years, AI researchers and practitioners from across IBM have trained advanced AI models and prototyped AI technologies on Vela, including IBM’s next-generation AI studio, watsonx.ai, which became generally available in July of 2023~\citep{granite_open}. Watsonx.ai is now deployed in many of IBM Cloud's global locations, a testament to the flexibility and global deployability of Vela's design. This section will provide examples of the outstanding performance we achieve on Vela without compromising on the flexible design requirement. 

In our training runs, tensor parallelism (TP), pipeline parallelism (PP) and data parallelism (DP) \citep{megatron1, megatron2, meagscale} are used to support the training of larger models on multiple GPUs to minimize the training time. For example, the Granite 20B Code model was trained on 768 GPUs with 4-way TP, 4-way PP, and 48-way DP. The point-to-point communication for PP and all-reduce for DP uses GDR communication channels, whereas the all-reduce for TP uses NVLink channels.

Table~\ref{granite-vela-models} shows a representative set of models trained on Vela over the past several months~\citep{granite_open}. The model sizes vary from 8B parameters to 20B parameters covering both language and code use cases. These models used a minimum of 768 GPUs, trained over a minimum of 2T tokens and trained for 30 or more days. A high performance Scale File system was used for input data access.

\begin{table}[!h]
    \begin{center}
    \begin{tabular}{ | c | c | c | c | c |}
    \hline
    Model &  GPUs &  Tokens & Training Duration & GPU hours \\ \hline\hline
    Granite-20B-code  & 768 & 2.1T & 46 days & 847872 \\\hline  
    Granite-13B-5LANG & 768 & 2.5T & 38 days & 718848 \\\hline  
    Granite-8B        & 1024 & 4T & 31 days & 768856 \\\hline 
    \end{tabular}
    \caption{A sample of the models trained on Vela. Each of the models is trained over 2T tokens and consumed over 700,000 GPU hours. Model architecture details can be found here~\citep{granite_open}.}
    \label{granite-vela-models}
    \end{center}
 \end{table}

To evaluate the competitiveness of this training throughput, we turn to two sources of information. First, we compare against models trained using the popular Megatron framework from the Megatron paper~\citep{megatron2} (see Table 1), which provides a reference of achieved TFLOPs per GPU for different model sizes. For model size we listed above between 8B and 20B, they report between 135 and 142 TFLOPs per GPU on an Infiniband system. While our model architectures are slightly different from the model described in that paper, we measured 140 TFLOPSs per GPU with our Granite-13B model running on 256 GPUs on Vela. Second, Bloomberg GPT~\citep{wu2023bloomberggpt} is trained on a cloud service provider with 512 A100 GPUs and they report 101TFLOPs per GPU; our internal experiments with the same model achieved 160TFLOPs per GPU on Vela. In addition, we evaluated our system with a large collection of test models from 3B to 75B and confirmed that the system achieves scalable performance across a range of GPU counts. 

In addition, we are pursuing a number of innovation vectors for further improving performance through various software and infrastructure optimizations, such as topology-based scheduling, tuning of the underlay network, optimization of the configurations of our network cards, and the use of different software frameworks such Pytorch native training with hybrid sharding, which can improve the overlap of compute with communication. Initial results from each of these future directions are showing promising results and represent new innovation vectors for optimizing the performance of workloads running on Vela. 

\subsection{Full picture of Vela technology stack}
\begin{figure}[!htb]
 \centering
  \fbox{\includegraphics[width=0.9\linewidth,trim={0.0cm 5cm 0cm 5cm},clip]{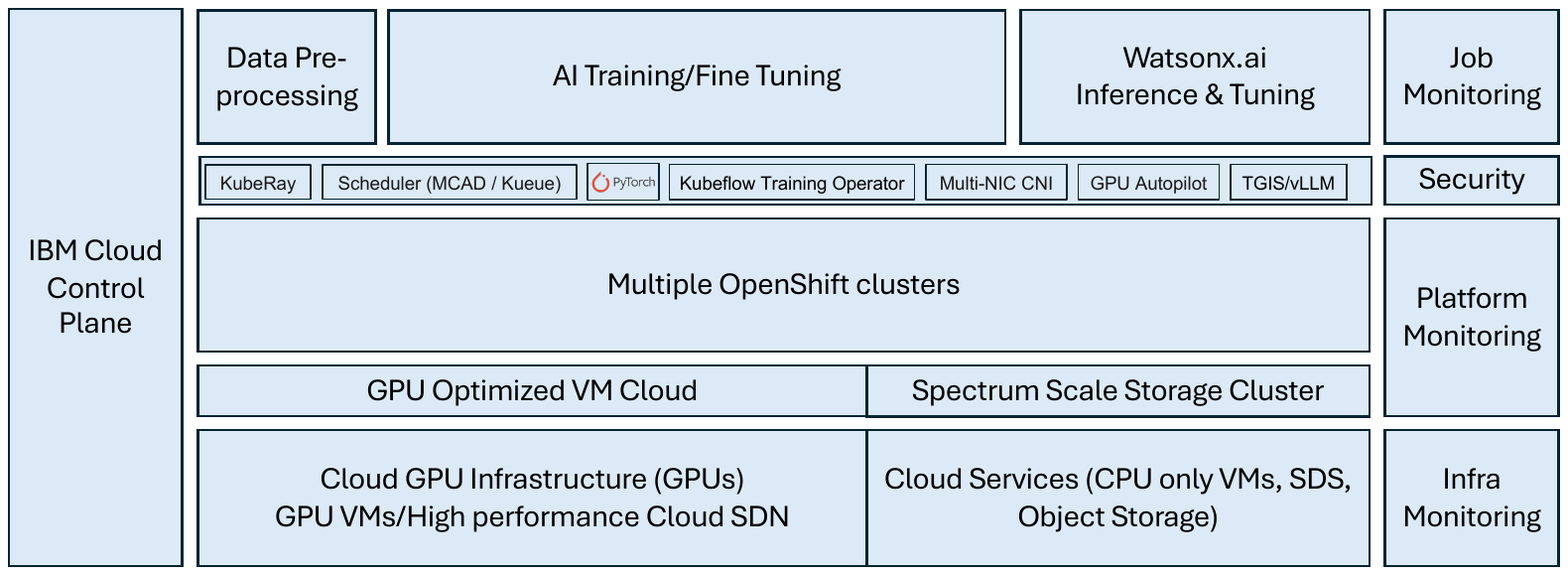}}
 \caption{Vela AI Cloud Reference Architecture Supporting End to End AI Life Cycle.}
       \label{ai-workflow}
 \end{figure}

So far, we discussed all the building blocks of the Vela system, including the cloud virtual machines, the storage system, the OpenShift container platform, etc. Figure~\ref{ai-workflow} summarizes how all of the technologies and innovations described in previous sections come together to enable high-performance AI compute on Vela, for use cases across the AI life-cycle. 

Our system administrators use IBM Cloud Control Plane APIs and subsets of the building blocks to construct distinct OpenShift clusters for data pre-processing, training and inference use cases. For example, the data pre-processing workloads tend to use mostly CPUs so those clusters will have many more CPU nodes. The training workloads require high-performance networking, a high-performance file system and GPUs so those cluster are built with those three technologies and the corresponding AI training stack. The watsonx.ai inference service tends to use 8 or fewer GPUs per model, so those clusters don't require high performance networking and storage. Inference services are also needed in many geographic locations so these clusters are built in multiple cloud data centers as well.

\clearpage

\parskip=1pt plus 1pt
\parindent=15pt
\section{Blue Vela AI Infrastructure}

In 2023, with the demand for AI compute continuing to grow exponentially, IBM initiated a new effort to build a world-class hosting environment dedicated entirely to large-scale AI model training. We began this effort by researching different options for compute, storage, and network infrastructure, evaluating various data centers and the facilities capabilities they offered, and, most importantly, determining what end-to-end architecture we would adopt for this platform. The following sections of this document detail the choices we made for the "Blue Vela" system, where we also provide implementation-related details.

\begin{figure}[h!]
 \centering
\begin{minipage}[t]{.99\linewidth}
      {\includegraphics[width=\linewidth]{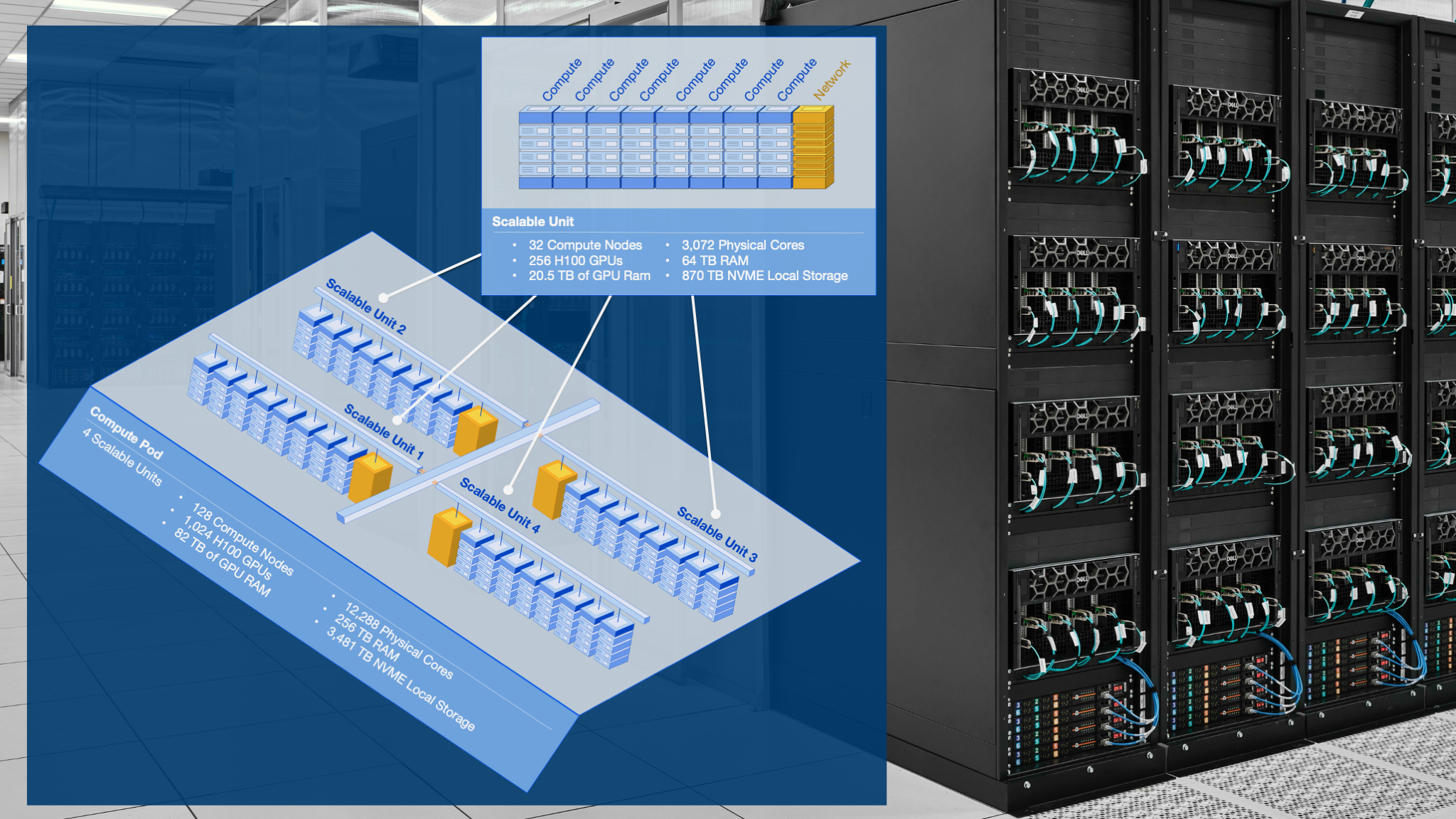}}%
  \end{minipage}%
  \caption
    {%
      Blue Vela building blocks and view of the data center%
      \label{fig:BlueVelaModularUnits}%
      }
 \end{figure}

The Blue Vela cluster, which began to come online in April 2024, has already significantly increased IBM Research's GPU capacity for training AI models, marking a 104\% increase over 2023. Once completed, Blue Vela will have provided an overall cumulative 214\% increase in available GPU capacity for AI training by the end of 2024 via the incremental roll-out of additional Compute Pods.  Figure \ref{fig:BlueVelaModularUnits} provides an overview of our building blocks of 32 nodes \textit{Scalable Units (SU)}, how four of these Scalable Units (SU) come together to form a 128 node \textit{Compute Pod} along with a glimpse of part of the system in the data center. While the Blue Vela system is comprised of a large number of power-intensive NVIDIA H100 GPUs, this cluster is hosted in a data center that utilizes 100\% renewable energy, minimizing our carbon footprint.

\subsection{Blue Vela Architecture}

\begin{figure}[H]
    \centering
    \includegraphics[width=0.95\linewidth]{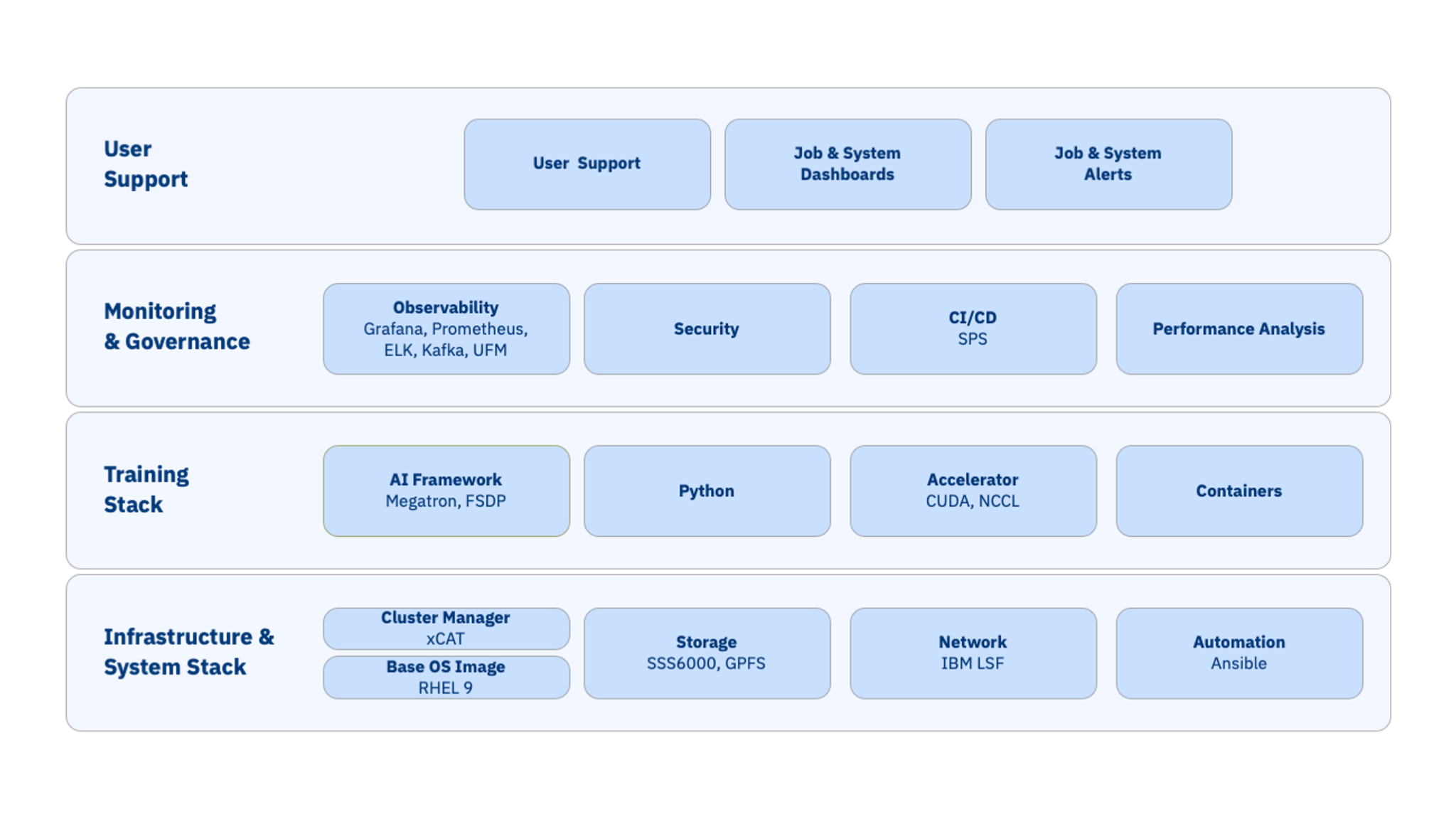}
    \caption{Layers of Blue Vela}
    \label{fig:BVLayers}
\end{figure}

Designed in partnership with Dell and NVIDIA, the Blue Vela Cluster is a state-of-the-art compute platform that is purpose-built to handle our most demanding model training tasks. Building on the NVIDIA H100 SuperPod reference architecture, we have customized Blue Vela to deliver high-performance GPU compute resources that best support our target workloads. The subsequent sections of this document will provide a comprehensive overview of our infrastructure, training stack, operational model, and workloads. Figure \ref{fig:BVLayers} provides a layered view of the Blue Vela infrastructure, system stack, training stack, monitoring \& governance, and user support layers.

\subsubsection{Network Infrastructure}

\begin{figure}[H]
   \centering
   \begin{minipage}[t]{0.85\linewidth}
     \subcaptionbox{Blue Vela Scalable Unit Infiniband Network}
       {\includegraphics[width=\linewidth]{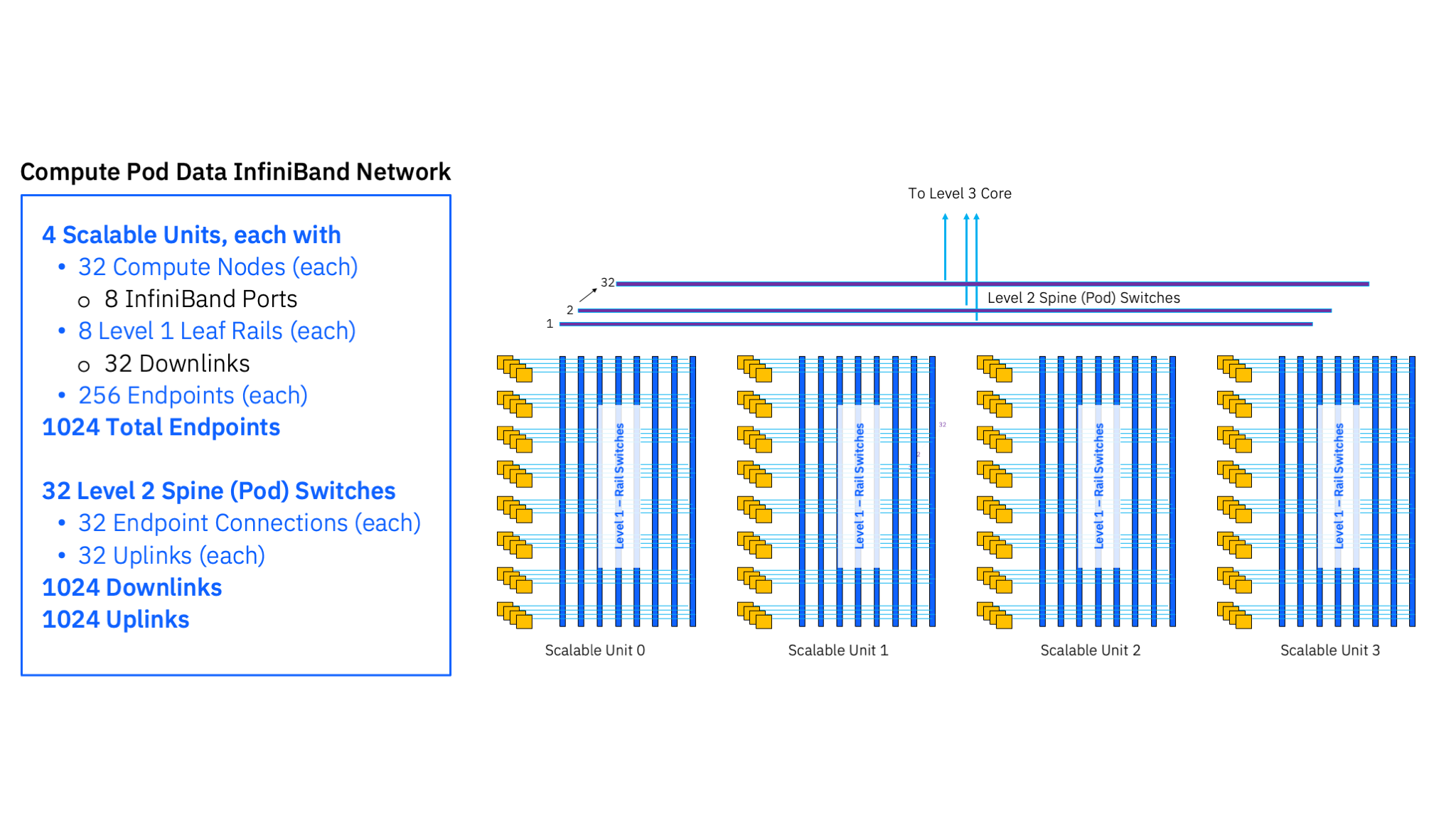}}%
   \end{minipage}%
   \hfill
   \begin{minipage}[b]{0.8\linewidth}
     \subcaptionbox{Blue Vela Compute Pod Infiniband Network}
       {\includegraphics[width=.95\linewidth]{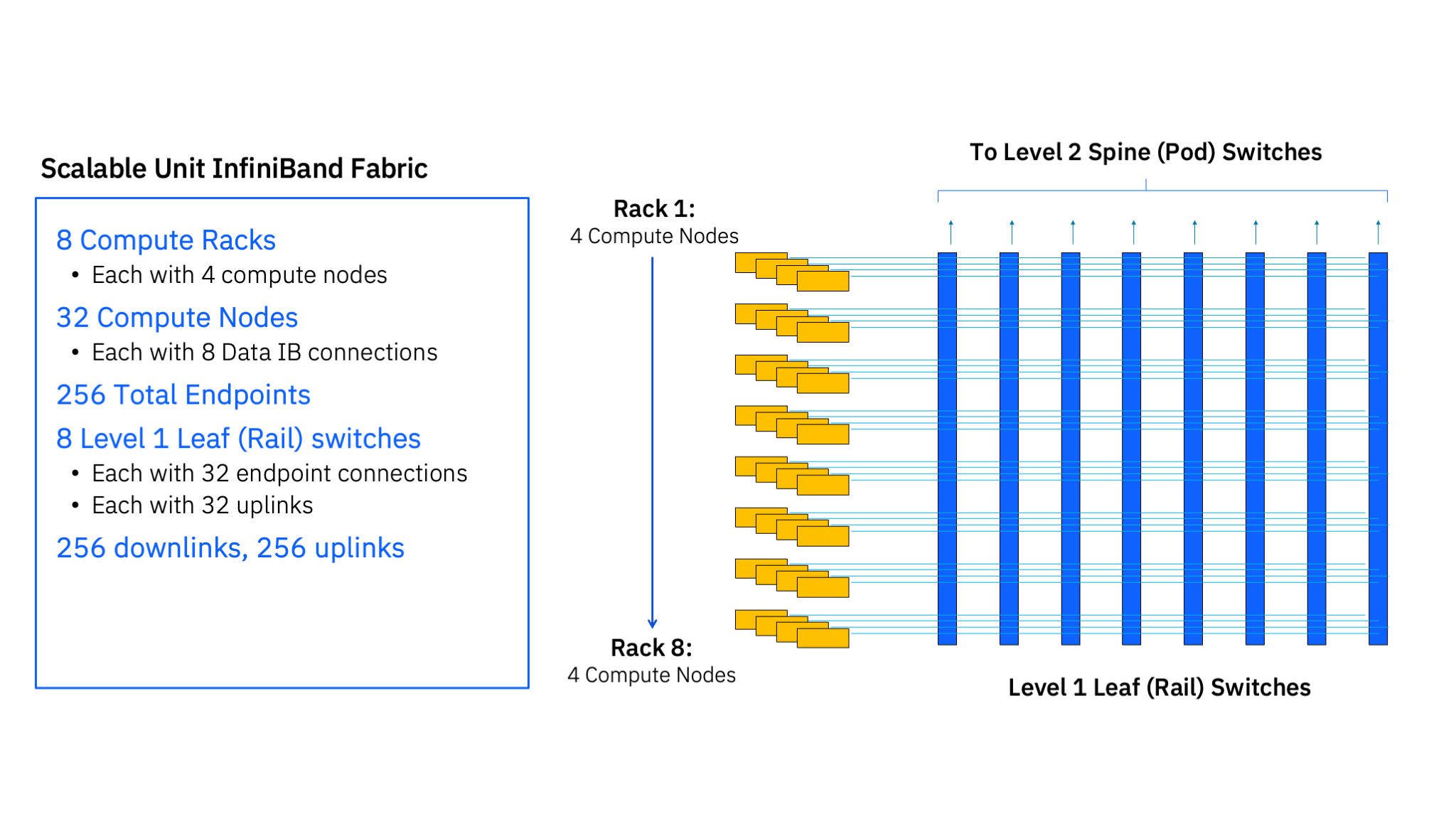}}
   \end{minipage}%
   \caption
     {%
       \label{fig:Metrics} Blue Vela Infiniband Network Architecture
     }%
 \end{figure}

As the number of GPUs utilized to train larger and more connected models increases, communication latency becomes a critical bottleneck. Therefore, the design of Blue Vela originated with the network. Blue Vela is designed around four distinct purpose-built networks. The first one is the compute InfiniBand fabric, which facilitates GPU-to-GPU communication, as shown in Figure~\ref{fig:Metrics}. The second is the storage InfiniBand fabric, which provides access to the storage subsystem, as described in the next section. The third network is the in-band Ethernet host network is used for inter-node communication outside the compute fabric. The fourth network is the out-of-band network, also called the management network, which provides access to the management interfaces on the servers and switches.

For over a decade, IBM Research has utilized InfiniBand networking in our on-prem AI Training clusters and, most notably in the Summit and Sierra Supercomputer Systems, deployed in 2018 at Oak Ridge (ORNL) and Lawrence Livermore (LLNL) Department of Energy (DOE) National Laboratories~\citep{summit2}. While the InfiniBand protocol may seem foreign and exotic compared to Ethernet, we have found that it provides peak performance while simultaneously being simpler to deploy and manage for high-performance on-premise clusters. Training today's large AI models imposes extreme demands on network fabrics, so we are utilizing a non-blocking Fat Tree architecture across all of our compute pods. This ensures that there are no oversubscribed links. Additionally, our network topology is rail-optimized, as recommended by NVIDIA, which further optimizes latency in all-reduce style operations. These operations are crucial for AI training workloads.

To ensure seamless scalability of our researchers’ applications across multiple GPUs, nodes, and compute pods, we have adopted NVIDIA’s reference architecture. This allows us to use their ecosystem of tools, libraries, and compilers for accelerated computing. Minimizing latency becomes crucial as the number of GPUs in your training job increases. To address this issue, we rely on NCCL, which is optimized to achieve high bandwidth and low latency. Additionally, it provides topology awareness to further improve performance in complex systems.

NVIDIA provides the Unified Fabric Manager (UFM) application to manage InfiniBand data center networks. Using dedicated UFMs to manage compute and storage fabrics has resulted in several management-related advantages on Blue Vela. As observed in Cloud Vela, large-scale clusters are susceptible and can quickly degrade if a single GPU is throttled or unavailable. However, with the help of UFM, these issues can be detected and addressed immediately on Blue Vela, thus ensuring that the GPUs are utilized to their maximum potential. Such a tool does not exist for Ethernet, which is why we had to build network monitoring on Cloud Vela. 

We also modified the standard storage fabric configuration to integrate IBM’s new Storage Scale System (SSS) 6000, which we were the first to deploy. We configured each server with two ConnectX-7 adapters, each capable of providing 400GbE, for a total of 800GbE per server. This setup allows us to accommodate future growth as the demand for training data increases as new modalities are explored.

Finally, our system has two dedicated Ethernet networks: a 100GbE In-band network and a 1G Out-of-Band network. The In-band network is used for non-GPU-GPU communication, specifically for monitoring and workload scheduling traffic. You can find more details about this in the software stack section. The out-of-band network is solely used for cluster management purposes and provides direct and secure access to server, switch, and iPDU management.

\subsubsection{Compute Infrastructure}
 
\paragraph{\textbf{Compute Nodes - }}
The starting point for selecting our optimal compute node configuration was rooted in the NVIDIA reference HGX Platform guidelines. Utilizing Dell's new PowerEdge XE9680, we adapted these recommendations based on historical workload data gathered from Vela and other on-premise AI Training clusters. The resulting Blue Vela compute node configuration is as follows:

\begin{itemize}[noitemsep,topsep=0pt]
    \item Dual 48-core 4th Gen Intel Xeon Scalable Processors 
    \item 2TB of RAM
    \item 8 NVIDIA H100 GPUs with 80GB High Bandwidth Memory (HBM)
    \item10 NVIDIA ConnectX-7 NDR 400 gigabits per second (Gb/s) InfiniBand Host Channel Adapters (HCA)
        \begin{itemize}[noitemsep,topsep=0pt]
        \item 8 dedicated to compute fabric
        \item 2 dedicated to storage fabric
    \end{itemize}
    \item 8 3.4TB Enterprise NVMe U.2 Gen4
    \item Dual 25G Ethernet Host links
    \item 1G Management Ethernet Port  \newline
\end{itemize}

\paragraph{\textbf{Management Nodes -}}
While the compute nodes were treated as homogenous and ephemeral resources, the management nodes are utilized to run key services such as authentication and authorization, workload scheduling, observability, and security. While these services have different performance characteristics and bottlenecks, we chose to have a standard, oversized hardware profile built on the Dell PowerEdge R760XS to simplify support and management and provide for future service expansion. The software stack section describes the hosted services in more detail. \\

The configuration of each management server is as follows:
\begin{itemize}[noitemsep,topsep=0pt]
    \item Dual 32-core 4th Gen Intel Xeon Scalable Processors 
    \item 1TB of RAM
    \item 2 NVIDIA® ConnectX-7 NDR 400 gigabits per second (Gb/s) InfiniBand Host Channel Adapters (HCA) dedicated to the storage fabric
    \item 2 1TB Enterprise NVMe U.2 Gen4
    \item 4 100G Ethernet Host links
    \item 1G Management Ethernet Port
\end{itemize}



\subsubsection{Storage}

Blue Vela was designed specifically to train large language models. The tasks of data crawling, deduplication, filtering, and tokenization of the training data occur in a cloud-based elastic environment~\citep{granite_open}, allowing us to prioritize storage tuning on reading tokenized datasets, writing model checkpoints, and general training data exhaust. 

Our storage subsystem was designed around the IBM Spectrum Scale~\citep{storage-scale} ecosystem and the new IBM Storage Scale System 6000~\citep{scale-ess6000}. Utilizing InfiniBand and PCIe Gen 5 technology for optimal performance, each SSS appliance is capable of delivering upwards of 310 GB/s throughput for reads and 155 GB/s for writes. While our InfiniBand storage fabric deployment allows us to deploy up to thirty-two SSS6000 appliances, we began with two fully populated SSS6000 chassis, with 48 30TB U.2G4 NVMes, which provides almost 3PB of raw storage. Each SSS appliance can also accommodate up to seven additional external JBOD enclosures, each up to 22TB, to expand capacity. IBM Storage Scale ecosystem and IBM Storage Scale System 6000 support automatic, transparent data caching to accelerate queries. 

\subsubsection{Data Center Selection and Design}

We considered three data center location options and conducted a feasibility and cost-benefit analysis based on the following criteria: site buildup and running cost, the feasibility of hosting the servers, and the timeline of implementation. Through this process, we identified an existing facility that could host the infrastructure. However, this site only had approximately half the required power and cooling capacity spread across three adjacent data center rooms. These constraints called for a creative problem-solving approach to make it work. A vital component of this approach was the creation of a Digital Twin, which our data center design team quickly assembled to explore and simulate the performance, financial, and environmental trade-offs to accommodate this project. Data center selection is a complex set of tradeoffs based on capacity utilization, risk management, and energy efficiency. This Digital Twin enabled us to efficiently explore various permutations of these factors to quickly mitigate the site's initial shortcomings.

\begin{figure}[!htb]
    \centering
    \includegraphics[width=\linewidth]{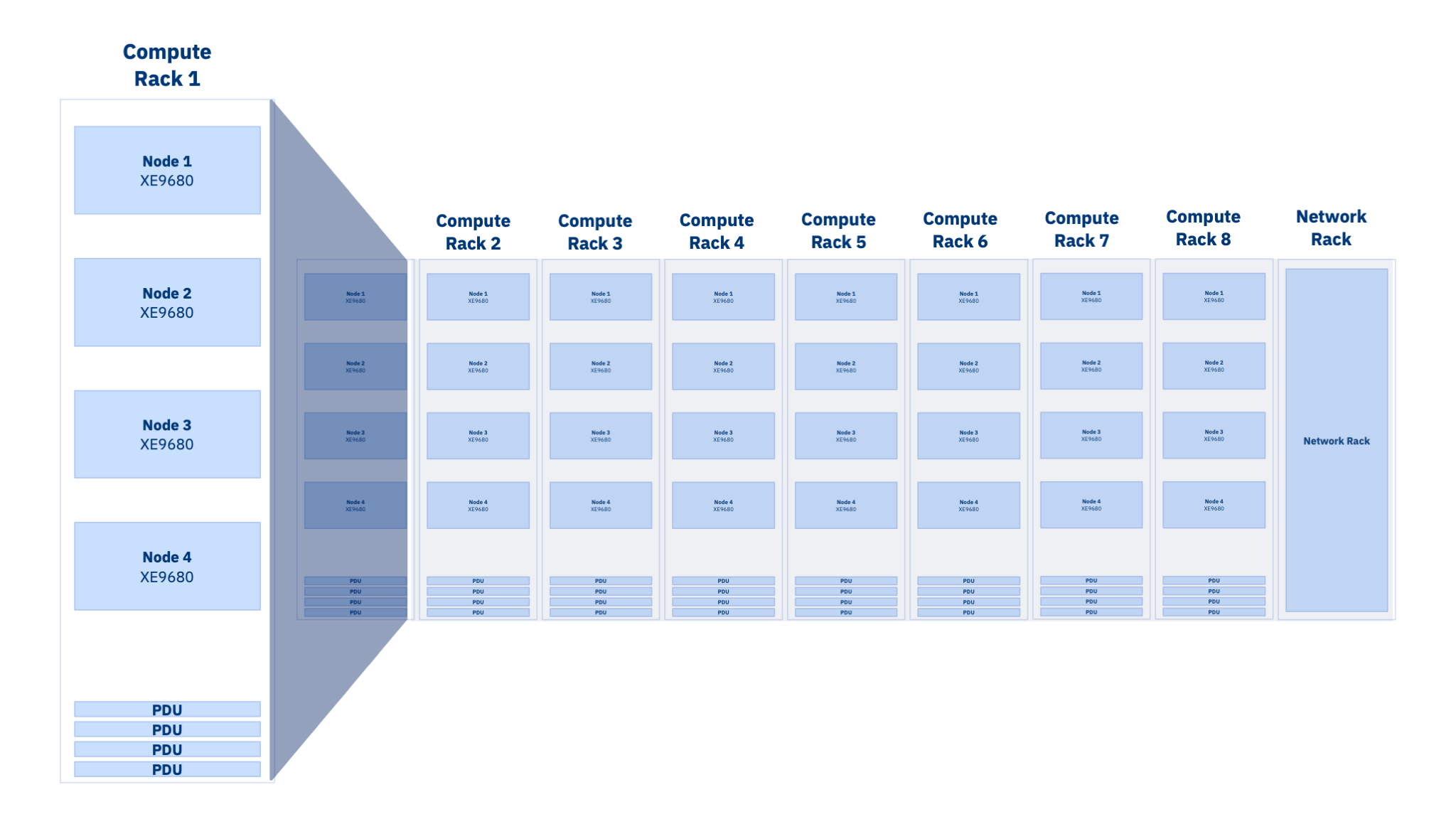}
    \caption{Blue Vela Compute Rack and Row Layout}
    \label{fig:BVRackRowLayout}
\end{figure}

While we have traditionally utilized rear-door heat exchangers to support a much greater GPU compute density than most data centers allow, the lack of water supply in this data center ruled out the implementation of this configuration. This forced us to design and implement an alternative means of achieving the density required to deliver the required GPU capacity while factoring in the 50-meter maximum length limitations of multi-mode fiber 400Gb IB cables. We utilized CAD modeling to design a solution that enabled us to maintain alignment with the reference architecture and include four compute nodes and four iPDUs per rack while remaining within a half-meter margin of cable length limitations. As part of this solution, we reconfigured rack locations and designed custom airflow containment enclosures that we interconnected via a custom signal cable raceway system. This allowed us to position the racks to utilize the existing airflow and heat rejection capability to accommodate a much heavier load than the data center was initially configured for. Figure \ref{fig:BVRackRowLayout} provides an overview of the layout of our compute node racks in relation to the standard row layout.

Since the rack power requirements and existing data center airflow capability meant that airflow recirculation/bypass had to be reduced to an extremely low level, we had to pay particular attention to sealing inside the racks. This included sealing all cable and power cord egress points, the underside of racks, rack-to-rack pass-throughs, and rack-to-containment infrastructure. We also had to define and place perforated floor tiles as determined by detailed Computational Fluid Dynamics (CFD) modeling for average to worst-case heat load and data center conditions.

Given the current power constraints, we reconfigured the facility's power level to bypass the UPS and primary power redundancy features to address power issues, increasing it by approximately 70\%. We also modified the Remote Power Panels (RPP) that supply power to the racks to meet our requirements. Finally, we developed a nine-month plan to incrementally increase the facility's power envelope to fully satisfy our needs. This includes complete facility-level redundancy and long-term uninterruptible power provided by the UPS units and generators. Therefore, power delivery is scheduled to catch up with the demand, and other UPS features will be retrofitted towards the end of the year. 

In addition to the modeling of the power, cooling, and layout of the cluster, our procurement team utilized the cluster modeling data to validate the purchase orders for the tens of thousands of cables required to connect a cluster of this size.

\subsection{Software Stack}

Given the complex nature of Blue Vela's compute, storage, and network components, the software stack was intentionally kept lean. With the ultimate objective of simplifying use, management, support, and troubleshooting, we selected a limited set of tools that addressed our critical needs. Our stack also needed to extend beyond what the Vela software stack addressed, supporting data center power, cooling, and system and network provisioning, which IBM Cloud provided on Vela.

\subsubsection{Host and Management Software Components}

We selected the eXtreme Cloud Automation Tool (xCAT) for system provisioning.
xCAT allows us to efficiently provision compute nodes using a stateless, in-memory operating system. This feature enables us to quickly reboot and re-image any problematic compute node and return them to a clean configuration within minutes. We use a traditional stateful OS for management nodes that host critical services. This setup balances the trade-off between critical services and quick reboot of non-critical components.

We chose Red Hat Enterprise Linux (RHEL) 9 as the host operating system based on previous experience and solid enterprise support. Using a full-featured Linux operating system on bare metal systems provides administrators with a variety of tools to diagnose and troubleshoot the hardware and software issues that arise with large-scale clusters.

To deploy and configure software across our development and production environments, we utilized Red Hat Ansible Automation Platform. Ansible provides a fast and repeatable process for software stack provisioning. 

\subsubsection{Workload Scheduling Software Components}

IBM Spectrum LSF is scheduling software that provides a resource management framework for Blue Vela. It evaluates users' job requirements, searches for the best resources to execute the job, and tracks its progress. LSF provides advanced workload management that features policy-driven scheduling, which optimizes the use of computing environments for HPC and AI workloads. While Blue Vela is a homogeneous computing environment, we selected LSF in part due to our long history of use on previous Research AI Training clusters, which has supported both x86 and PowerPC hardware as well as 7 generations of NVIDIA GPUs and is still extensively used today.  

The robust architecture of LSF is designed with fault tolerance in mind. Every component in the system has a recovery operation so that vital components are monitored by another component and can automatically recover from a failure. LSF continues to operate even if some of the hosts in the cluster are unavailable. One host in the cluster acts as the management host, but if the management host becomes unavailable, another management host candidate will take over. LSF can tolerate the failure of any host or group of hosts in the cluster. When a host becomes unavailable, all jobs that are running on that host are either requeued or lost, depending on whether the job was marked as rerunnable. No other pending or running jobs are affected.

LSF also provides deep integration with NVIDIA GPUs, allowing the scheduler to be aware of GPU status and utilization, recognize common GPU hardware issues, such as NVLink and ECC Memory errors, and take those into account in scheduling decisions.

\subsubsection{Observability Software Components}

Observability in the Blue Vela cluster differs from the Vela cluster. Unlike Vela, which is hosted on the IBM Cloud and can utilize its observability stack, the Blue Vela cluster is hosted in an on-premises data center. This means that IBM Research has ownership and responsibility for all solution components, from the infrastructure layer to the software stack. The upcoming sections will cover Blue Vela observability in this context.

Due to the size and scale of the cluster, the individual parts exhibit interesting and often unintentional interactions. This complexity and the excessive cost of downtime put significant pressure on the observability layer. For these reasons, we utilize the standard Elastic, Logstash, and Kibana (ELK) stack, along with Kafka for log aggregation and the Prometheus, Thanos, and Grafana stack for telemetry. 

The ELK stack is integrated tightly with LSF via IBM Spectrum LSF Explorer, allowing our business and technical users to rapidly create and view reports and dashboards to understand cluster utilization. We reinforce this data with additional structured data from system and service logs to correlate issues, improve anomaly detection, and provide an API to system information so we can improve telemetry. In addition, we utilize Kafka to ship sensitive information to our centralized ELK cluster, which is responsible for collecting data required by internal audit processes. 

Prometheus provides the basis for our telemetry platform. Deployed in an High Availability (HA) pair, we utilize a combination of open-source and custom exporters to collect valuable data into the Prometheus instances for use by administrators, users, and executives. Grafana provides a single landing page for cluster telemetry, incorporating host and GPU information from the compute nodes and service information from ELK and custom exporters. 

Due to the large volume of data produced by the Blue Vela system, we keep 30 days' worth of data in our local Prometheus servers. We deploy Thanos in an OpenShift Cluster hosted in IBM Cloud to gather and store historical data. Thanos, by default, leverages the Prometheus 2.0 storage format to store historical metric data in object storage while retaining fast query latencies. Additionally, Thanos provides a global query view across multiple Prometheus installations, merging data from our Prometheus HA pairs on the fly, allowing for maintenance windows without losing data, and efficiently downsampling data over time to optimize long-term storage costs. 





\subsection{Operational Model} 
The Granite models \citep{granite_open}, like other Foundational Models, differ from conventional AI workflows. The latter involves training small models that use GPU resources for short durations. In contrast, Granite models can vary in sizes and require significant GPU resources for extended periods of time. To meet our primary objectives of supporting the GPU compute capacity needs of Granite model training and maximizing the efficient use of this capacity, we developed an operational model that minimizes the impact of environmental, platform, and system-level issues on long-running model training jobs. 

This operational model requires end-to-end monitoring of all layers of our solution, along with predictive analysis, to estimate the duration of jobs and to enable the detection of any anomalies that may occur. Another key element of this operational model is automation, which executes runbook-driven system recovery when needed to enable the rapid restart of training jobs. Automation also plays a significant role in ensuring ongoing compliance and consistency across all environments.  Additionally, a robust change management process is required to eliminate the chances of any catastrophic changes being introduced into the environment. The cluster also utilizes the IBM Secure Pipelines Service (SPS) - a pipeline that, in combination with the Toolkit, helps ensure compliance and build standard environments to improve reproducibility. Any subsequent changes to the environment are managed through a robust change management process to eliminate any catastrophic changes to the cluster.
In addition, a comprehensive set of dashboards targeting specific personas, such as researchers, system admins, and executives, is required to enable the visualization of the vast range of monitoring data we collect in the appropriate context. Finally, we have the AskETE chatbot, which is our watsonx-powered chatbot trained on historical support content and cluster documentation. This chatbot provides automated support to researchers and enables the prompt ticketing of issues, freeing up human resources for other support tasks.

\subsection{Monitoring} To support large-scale AI model training workloads, such as Granite model-related training jobs, and minimize job downtime, our dashboards have been configured to collect a comprehensive set of 1180 metrics. Table \ref{tab: metric-collection}, below, details our time-based metric collection intervals.
\begin{table}[!h]
    \begin{center}
    \begin{tabular}{ | c | c |}
    \hline
        Time Frame & Interval\\ \hline\hline
        0-30 Days & All Data (based on capture interval)\\\hline
        30-90 Days & 5 Minute\\\hline
        90-365 Days & 1 Hour\\\hline
    \end{tabular}
    \caption{Metric collection details.}
    \label{tab: metric-collection}
    \end{center}
 \end{table}

GPU metrics sourced from NVIDIA Data Center GPU Manager(DCGM) are reported at 5 second intervals. System-level metrics were recorded at 60 second intervals. These two sources enable near real-time monitoring of the cluster.

\begin{figure}[!htb]
    \centering
    \includegraphics[width=0.95\linewidth]{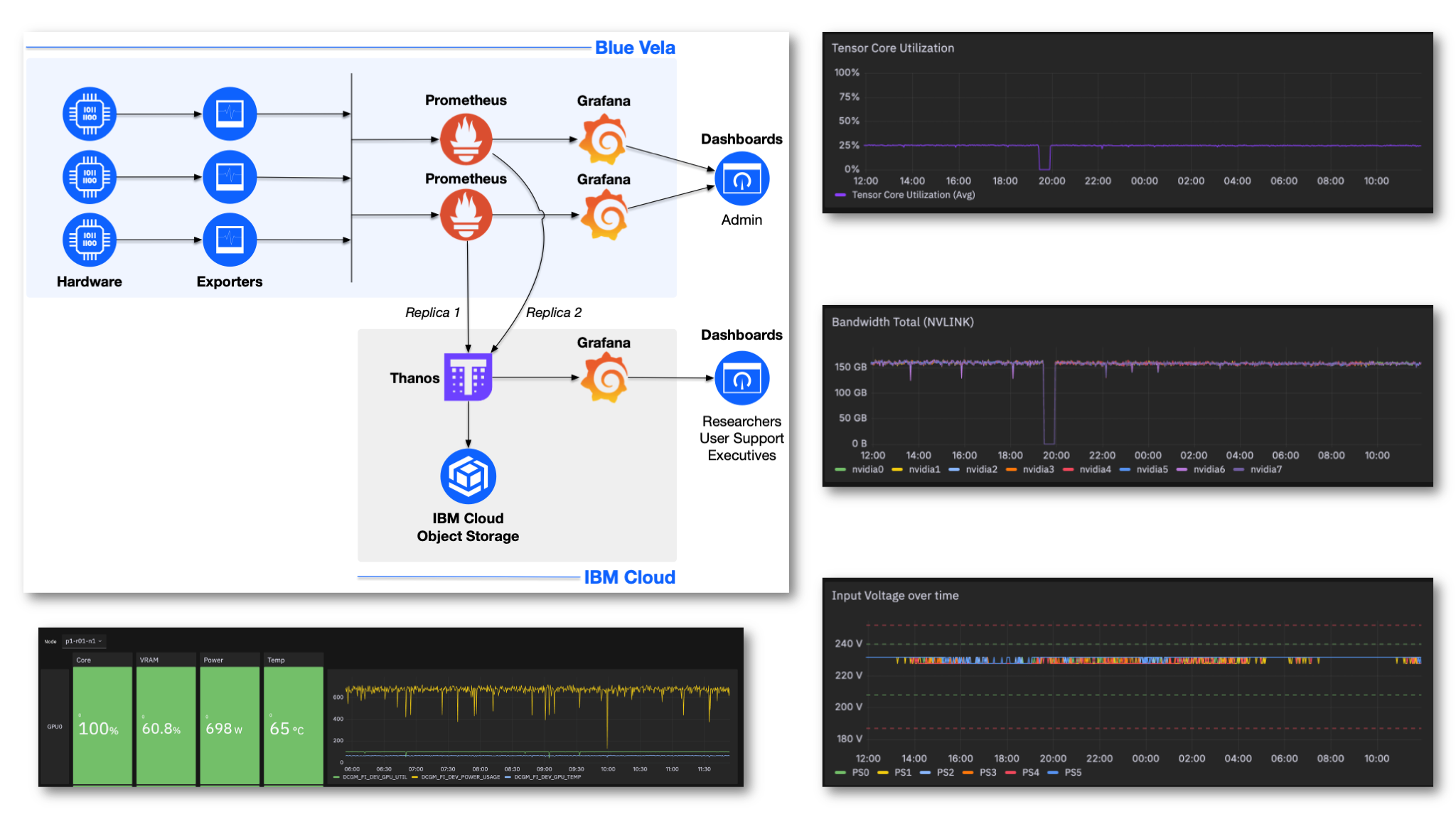}
    \caption{Dashboard Architecture \& GPU Monitoring Dashboards}
    \label{fig:Monitoring}
\end{figure}

These integrated data points and metrics also support full system observability (as described in the Software Stack | Observability section) , enabling ongoing optimization, diagnosis, and system characterization (Figure \ref{fig:Monitoring}), for example:
\begin{itemize}[noitemsep,topsep=0pt]
\item \textit{GPU Tensor core utility}: Tensor core utility is a key indicator of how a job is performing and can indicate if a job is performing and well optimized.
\item \textit{GPU overall utility}: If GPU overall utility drops it helps indicate that the node may be underperforming.
\item \textit{System health}: monitors the physical health or the system such as GPU health, and memory health. These metrics help to monitor whether the systems are healthy and run as smoothly as possible since if a single node in a job is unhealthy, the entire job performs as fast as its slowest node due to the bottleneck. 
\item \textit{Power usage from the IPDUs to individual GPU}: These metrics are useful to see how much power the entire data center is drawing as well as seeing if 1 node is underperforming, such as when a single node or a single GPU is drawing less power then others which may indicate an issue
\item \textit{LSF status}: this metric indicates if a node is available and open to jobs vs if a node is closed and will not accept any jobs.
\item \textit{GPU throttling and its reason}: this will show if a GPU is not performing at 100\% and is useful to identify thermal issues and GPU slows down. Under these scenarios, the systems can get too hot or a power break slowdown occurs.
\end{itemize}


\subsection{Initial Workload performance on Blue Vela}


Blue Vela has undergone a rapid bring-up and immediately made an impact on model training. Within the first month of production use (beginning April 1st, 2024), the first set of models has been trained and open-sourced (May 6th, 2024) - which is a major milestone for the granite model family \citep{granite_open}. From the onset, the infrastructure has demonstrated good potential in throughput and has already shown a $5\%$ higher performance out-of-the-box compared to other environments of the same configuration. While we are in the midst of delivery and working to tune the system, this section highlights our current preliminary results of the initial jobs and their performance. The current performance of the cluster shows high throughputs (90-321B per day depending on the training setting and the model being trained) (Table ~\ref{tab: bluevela-training}). There is a large potential to further optimize the performance and throughputs once all the pod integration has been completed. With Operator fusion and FSDP, the throughputs are expected to improve by another 25-30\%.

\begin{table}[!h]
    \begin{center}
        \caption{Blue Vela preliminary results.}
        \label{tab: bluevela-training}
    \begin{tabular}{ | c | c | c |}
    \hline
    Model & Number of GPUs & Number of Tokens per day  \\ \hline\hline
    dolomite-FSDP & 256 & 321B  \\\hline  
    Granite-34B  & 448 & 47B  \\\hline 
    Granite-20B (Megatron) & 512 & 94.8B \\\hline  
    Granite 34B (Megatron)   & 1024 & 90.8B  \\\hline 
    \end{tabular}
    \end{center}
 \end{table}





\subsection{Future Directions for Blue Vela Development}

We launched the Blue Vela initiative with a specific goal: to provide researchers with sufficient GPU capacity to meet their AI model training needs as soon as possible. To achieve this objective, we planned to deploy a vast range of resources in the shortest possible time while ensuring their efficient and effective utilization. Unlike Vela, which relied heavily on innovation across all solution layers, we intended to use a proven, high-performing, industry-standard reference architecture and corresponding infrastructure components. While we welcomed opportunities to innovate, they were limited to where absolutely necessary, primarily where it was required to support the integration of custom infrastructure components or operational processes. The launch of the Blue Vela cluster represents the accomplishment of our primary objective: a major expansion of GPU compute capacity required to support IBM Research’s AI model training needs. To date, Blue Vela is actively being utilized to run Granite model training jobs. 

With our end-to-end responsibility for all layers of this solution, including operations, we will continue to improve upon its usage experience and drive even higher levels of efficient and effective utilization. Moving forward, we will continue to focus on system and software optimization to accelerate AI model training, the automation of job scheduling and job restarts, and using AIOps for anomaly detection in cluster operational data.

\section{Summary}

This document detailed IBM’s hybrid-cloud-based approach to building world-class infrastructure to support IBM’s model development activities at scale. We have built two state-of-the-art environments: (1) An AI-optimized supercomputer natively integrated into the IBM Cloud, that provides state-of-the-art performance, scalability, multi-tenancy and flexible deployability across IBM Cloud’s global footprint and (2) An AI-optimized on-premise supercomputer to support our largest and most ambitious AI model training tasks. Taken together, they provide IBM with the ability to train industry-leading models, bring them to our clients, maintain business agility, and future-proof us against the rapidly evolving model landscape in the industry. IBM will continue advancing our AI infrastructure capabilities, both in the IBM Cloud and on-premise, in line with our overall hybrid cloud and AI strategy.

\addcontentsline{toc}{section}{Bibliography}
\bibliography{main}

\begin{thebibliography}{33}
\providecommand{\natexlab}[1]{#1}
\providecommand{\url}[1]{\texttt{#1}}
\expandafter\ifx\csname urlstyle\endcsname\relax
  \providecommand{\doi}[1]{doi: #1}\else
  \providecommand{\doi}{doi: \begingroup \urlstyle{rm}\Url}\fi

\bibitem[ibm({\natexlab{a}})]{ibm-mcad}
Multi-cluster app dispatcher, {\natexlab{a}}.
\newblock URL
  \url{https://github.com/project-codeflare/multi-cluster-app-dispatcher}.

\bibitem[ibm({\natexlab{b}})]{ibm-watsonx}
Ibm watsonx, {\natexlab{b}}.
\newblock URL \url{https://www.ibm.com/artificial-intelligence}.

\bibitem[lla()]{llama3}
Introducing meta llama 3: The most capable openly available llm to date.
\newblock URL \url{https://ai.meta.com/blog/meta-llama-3/}.

\bibitem[met()]{meta-ai-clusters}
Building meta’s genai infrastructure.
\newblock URL
  \url{https://engineering.fb.com/2024/03/12/data-center-engineering/building-metas-genai-infrastructure/}.

\bibitem[Gershon et~al.(2023{\natexlab{a}})Gershon, Karacali-Akyamac, Seelam,
  Thorstensen, and Badlaney]{vela-upgrade}
Talia Gershon, Bengi Karacali-Akyamac, Seetharami Seelam, Drew Thorstensen, and
  Rohit Badlaney.
\newblock Supercharging ibm’s cloud-native ai supercomputer,
  2023{\natexlab{a}}.
\newblock URL
  \url{https://research.ibm.com/blog/vela-ai-supercomputer-updates}.

\bibitem[Gershon et~al.(2023{\natexlab{b}})Gershon, Seelam, Jubran, Gampel, and
  Thorstensen]{vela-intro}
Talia Gershon, Seetharami Seelam, Jay Jubran, Eran Gampel, and Drew
  Thorstensen.
\newblock Why we built an ai supercomputer in the cloud, 2023{\natexlab{b}}.
\newblock URL
  \url{https://research.ibm.com/blog/AI-supercomputer-Vela-GPU-cluster}.

\bibitem[Jacob et~al.(2019)Jacob, Ming-Wei, Kenton, and
  Kristina]{devlin2019bert}
Devlin Jacob, Chang Ming-Wei, Lee Kenton, and Toutanova Kristina.
\newblock Bert: Pre-training of deep bidirectional transformers for language
  understanding, 2019.

\bibitem[Jiang et~al.(2024)Jiang, Lin, Zhong, Huang, Chen, Zhang, Peng, Li,
  Xie, Nong, Jia, He, Chen, Bai, Hou, Yan, Zhou, Sheng, Jiang, Xu, Wei, Zhang,
  Nie, Zou, Zhao, Xiang, Liu, Li, Jia, Ye, Jin, and Liu]{meagscale}
Ziheng Jiang, Haibin Lin, Yinmin Zhong, Qi~Huang, Yangrui Chen, Zhi Zhang,
  Yanghua Peng, Xiang Li, Cong Xie, Shibiao Nong, Yulu Jia, Sun He, Hongmin
  Chen, Zhihao Bai, Qi~Hou, Shipeng Yan, Ding Zhou, Yiyao Sheng, Zhuo Jiang,
  Haohan Xu, Haoran Wei, Zhang Zhang, Pengfei Nie, Leqi Zou, Sida Zhao, Liang
  Xiang, Zherui Liu, Zhe Li, Xiaoying Jia, Jianxi Ye, Xin Jin, and Xin Liu.
\newblock {MegaScale}: Scaling large language model training to more than
  10,000 {GPUs}.
\newblock In \emph{21st USENIX Symposium on Networked Systems Design and
  Implementation (NSDI 24)}, pp.\  745--760, Santa Clara, CA, April 2024.
  USENIX Association.
\newblock ISBN 978-1-939133-39-7.
\newblock URL
  \url{https://www.usenix.org/conference/nsdi24/presentation/jiang-ziheng}.

\bibitem[Lockwood et~al.(2014)Lockwood, Tatineni, and Wagner]{lockwood2014sr}
Glenn~K Lockwood, Mahidhar Tatineni, and Rick Wagner.
\newblock Sr-iov: Performance benefits for virtualized interconnects.
\newblock In \emph{Proceedings of the 2014 Annual Conference on Extreme Science
  and Engineering Discovery Environment}, pp.\  1--7, 2014.

\bibitem[Mishra et~al.(2024)Mishra, Stallone, Zhang, Shen, Prasad, Soria,
  Merler, Selvam, Surendran, Singh, Sethi, Dang, Li, Wu, Zawad, Coleman, White,
  Lewis, Pavuluri, Koyfman, Lublinsky, de~Bayser, Abdelaziz, Basu, Agarwal,
  Zhou, Johnson, Goyal, Patel, Shah, Zerfos, Ludwig, Munawar, Crouse,
  Kapanipathi, Belgodere, Salaria, Calio, Wen, Seelam, Fonseca, Singhee, Desai,
  Cox, Puri, and Panda]{granite_open}
Mayank Mishra, Matt Stallone, Gaoyuan Zhang, Yikang Shen, Aditya Prasad,
  Adriana~Meza Soria, Michele Merler, Parameswaran Selvam, Saptha Surendran,
  Shivdeep Singh, Manish Sethi, Xuan-Hong Dang, Pengyuan Li, Kun-Lung Wu, Syed
  Zawad, Andrew Coleman, Matthew White, Mark Lewis, Raju Pavuluri, Yan Koyfman,
  Boris Lublinsky, Maximilien de~Bayser, Ibrahim Abdelaziz, Kinjal Basu, Mayank
  Agarwal, Yi~Zhou, Chris Johnson, Aanchal Goyal, Hima Patel, Yousaf Shah,
  Petros Zerfos, Heiko Ludwig, Asim Munawar, Maxwell Crouse, Pavan Kapanipathi,
  Brian Belgodere, Shweta Salaria, Bob Calio, Sophia Wen, Seetharami Seelam,
  Carlos Fonseca, Amith Singhee, Nirmit Desai, David~D. Cox, Ruchir Puri, and
  Rameswar Panda.
\newblock Granite code models: A family of open foundation models for code
  intelligence.
\newblock 2024.
\newblock URL \url{https://arxiv.org/abs/2405.04324}.

\bibitem[Mohan \& Sheard(2022)Mohan and Sheard]{gtc22}
Apoorve Mohan and Matthew Sheard.
\newblock How to deploy a high-performance distributed ai training cluster with
  nvidia gpus and kvm, 2022.
\newblock URL
  \url{https://www.nvidia.com/en-us/on-demand/session/gtcspring22-s42633/}.

\bibitem[Narayanan et~al.(2021)Narayanan, Shoeybi, Casper, et~al.]{megatron2}
Deepak Narayanan, Mohammad Shoeybi, Jared Casper, et~al.
\newblock Efficient large-scale language model training on gpu clusters using
  megatron-lm.
\newblock In \emph{Proceedings of the International Conference for High
  Performance Computing, Networking, Storage and Analysis}, SC '21, New York,
  NY, USA, 2021. Association for Computing Machinery.
\newblock ISBN 9781450384421.
\newblock \doi{10.1145/3458817.3476209}.
\newblock URL \url{https://doi.org/10.1145/3458817.3476209}.

\bibitem[NVIDIA(2023)]{cudasamples}
NVIDIA.
\newblock Nvidia cuda samples, 2023.
\newblock URL \url{https://github.com/NVIDIA/cuda-samples}.

\bibitem[NVIDIA(2024)]{hbm-row-remap}
NVIDIA.
\newblock Nvidia gpu memory error management, 2024.
\newblock URL
  \url{https://docs.nvidia.com/deploy/pdf/a100-gpu-mem-error-mgmt.pdf}.

\bibitem[Ott et~al.(2018)Ott, Edunov, Grangier, and Auli]{ott2018scaling}
Myle Ott, Sergey Edunov, David Grangier, and Michael Auli.
\newblock Scaling neural machine translation, 2018.

\bibitem[Raffel et~al.(2020)Raffel, Shazeer, Roberts, Lee, Narang, Matena,
  Zhou, Li, and Liu]{raffel2020exploring}
Colin Raffel, Noam Shazeer, Adam Roberts, Katherine Lee, Sharan Narang, Michael
  Matena, Yanqi Zhou, Wei Li, and Peter~J. Liu.
\newblock Exploring the limits of transfer learning with a unified text-to-text
  transformer, 2020.

\bibitem[Research(2024{\natexlab{a}})]{autopilot}
IBM Research.
\newblock Ai training autopilot, 2024{\natexlab{a}}.
\newblock URL \url{https://github.com/IBM/autopilot}.

\bibitem[Research(2024{\natexlab{b}})]{multinic-cni}
IBM Research.
\newblock Multi-nic cni, 2024{\natexlab{b}}.
\newblock URL \url{https://github.com/foundation-model-stack/multi-nic-cni}.

\bibitem[Schmuck \& Haskin(2002)Schmuck and Haskin]{storage-gpfs}
Frank Schmuck and Roger Haskin.
\newblock Gpfs: A shared-disk file system for large computing clusters.
\newblock In \emph{Conference on file and storage technologies (FAST 02)},
  2002.

\bibitem[Seelam et~al.(2023)Seelam, Mohan, Chen, and Chung]{vmvsbm}
Seetharami Seelam, Apoorve Mohan, Ming-Hung Chen, and IHsin Chung.
\newblock To virtualize or not to virtualize ai infrastructure: A perspective.
\newblock \emph{In ISCA workshop on HotInfr}, 2023.
\newblock URL \url{https://hotinfra23.github.io/papers/hotinfra23-paper16.pdf}.

\bibitem[Shoeybi et~al.(2019)Shoeybi, Patwary, Puri, et~al.]{megatron1}
Mohammad Shoeybi, Mostofa Patwary, Raul Puri, et~al.
\newblock Megatron-lm: Training multi-billion parameter language models using
  model parallelism.
\newblock \emph{arXiv preprint arXiv:1909.08053}, 2019.

\bibitem[Shoeybi et~al.(2020)Shoeybi, Patwary, Puri, LeGresley, Casper, and
  Catanzaro]{shoeybi2020megatronlm}
Mohammad Shoeybi, Mostofa Patwary, Raul Puri, Patrick LeGresley, Jared Casper,
  and Bryan Catanzaro.
\newblock Megatron-lm: Training multi-billion parameter language models using
  model parallelism, 2020.

\bibitem[Storage(2024{\natexlab{a}})]{scale-afm}
IBM Storage.
\newblock Active file management, 2024{\natexlab{a}}.
\newblock URL
  \url{https://www.ibm.com/docs/en/storage-scale/5.1.9?topic=overview-active-file-management}.

\bibitem[Storage(2024{\natexlab{b}})]{scale-cnsa}
IBM Storage.
\newblock Ibm storage scale container native documentation, 2024{\natexlab{b}}.
\newblock URL \url{https://www.ibm.com/docs/en/scalecontainernative}.

\bibitem[Storage(2024{\natexlab{c}})]{scale-cnsa-remote-mount}
IBM Storage.
\newblock Performing installation for a remote ibm spectrum scale cluster
  mount, 2024{\natexlab{c}}.
\newblock URL
  \url{https://www.ibm.com/docs/en/stgenablercontainers/2.1.0?topic=installation-performing-remote-spectrum-scale-cluster-mount}.

\bibitem[Storage(2024{\natexlab{d}})]{scale-ess6000}
IBM Storage.
\newblock Ibm storage scale system 6000, 2024{\natexlab{d}}.
\newblock URL \url{https://www.ibm.com/downloads/cas/JBVQYVXB}.

\bibitem[Storage(2024{\natexlab{e}})]{scale-fileset-based-volumes}
IBM Storage.
\newblock Storage class for creating fileset-based volumes, 2024{\natexlab{e}}.
\newblock URL
  \url{https://www.ibm.com/docs/en/scalecsi/2.10?topic=class-storage-creating-fileset-based-volumes}.

\bibitem[Storage(2024{\natexlab{f}})]{storage-scale}
IBM Storage.
\newblock Ibm storage scale documentation, 2024{\natexlab{f}}.
\newblock URL \url{https://www.ibm.com/docs/en/storage-scale}.

\bibitem[Stunkel et~al.(2020)Stunkel, Graham, Shainer, Kagan, Sharkawi,
  Rosenburg, and Chochia]{summit2}
C.~B. Stunkel, R.~L. Graham, G.~Shainer, M.~Kagan, S.~S. Sharkawi,
  B.~Rosenburg, and G.~A. Chochia.
\newblock The high-speed networks of the summit and sierra supercomputers.
\newblock \emph{IBM Journal of Research and Development}, 64\penalty0
  (3/4):\penalty0 3:1--3:10, 2020.
\newblock \doi{10.1147/JRD.2020.2967330}.

\bibitem[Touvron et~al.(2023)Touvron, Martin, Stone, Albert, Almahairi, Babaei,
  Bashlykov, Batra, Bhargava, Bhosale, Bikel, Blecher, Ferrer, Chen, Cucurull,
  Esiobu, Fernandes, Fu, Fu, Fuller, Gao, Goswami, Goyal, Hartshorn, Hosseini,
  Hou, Inan, Kardas, Kerkez, Khabsa, Kloumann, Korenev, Koura, Lachaux, Lavril,
  Lee, Liskovich, Lu, Mao, Martinet, Mihaylov, Mishra, Molybog, Nie, Poulton,
  Reizenstein, Rungta, Saladi, Schelten, Silva, Smith, Subramanian, Tan, Tang,
  Taylor, Williams, Kuan, Xu, Yan, Zarov, Zhang, Fan, Kambadur, Narang,
  Rodriguez, Stojnic, Edunov, and Scialom]{touvron2023llama2openfoundation}
Hugo Touvron, Louis Martin, Kevin Stone, Peter Albert, Amjad Almahairi, Yasmine
  Babaei, Nikolay Bashlykov, Soumya Batra, Prajjwal Bhargava, Shruti Bhosale,
  Dan Bikel, Lukas Blecher, Cristian~Canton Ferrer, Moya Chen, Guillem
  Cucurull, David Esiobu, Jude Fernandes, Jeremy Fu, Wenyin Fu, Brian Fuller,
  Cynthia Gao, Vedanuj Goswami, Naman Goyal, Anthony Hartshorn, Saghar
  Hosseini, Rui Hou, Hakan Inan, Marcin Kardas, Viktor Kerkez, Madian Khabsa,
  Isabel Kloumann, Artem Korenev, Punit~Singh Koura, Marie-Anne Lachaux,
  Thibaut Lavril, Jenya Lee, Diana Liskovich, Yinghai Lu, Yuning Mao, Xavier
  Martinet, Todor Mihaylov, Pushkar Mishra, Igor Molybog, Yixin Nie, Andrew
  Poulton, Jeremy Reizenstein, Rashi Rungta, Kalyan Saladi, Alan Schelten, Ruan
  Silva, Eric~Michael Smith, Ranjan Subramanian, Xiaoqing~Ellen Tan, Binh Tang,
  Ross Taylor, Adina Williams, Jian~Xiang Kuan, Puxin Xu, Zheng Yan, Iliyan
  Zarov, Yuchen Zhang, Angela Fan, Melanie Kambadur, Sharan Narang, Aurelien
  Rodriguez, Robert Stojnic, Sergey Edunov, and Thomas Scialom.
\newblock Llama 2: Open foundation and fine-tuned chat models, 2023.
\newblock URL \url{https://arxiv.org/abs/2307.09288}.

\bibitem[Walkup et~al.(2022)Walkup, Seelam, and Wen]{walkup2022best}
Robert Walkup, Seetharami~R Seelam, and Sophia Wen.
\newblock Best practices for hpc workloads on public cloud platforms: A guide
  for computational scientists to use public cloud for hpc workloads.
\newblock In \emph{Proceedings of the 2022 ACM/SPEC on International Conference
  on Performance Engineering}, pp.\  29--35, 2022.

\bibitem[Wu et~al.(2023)Wu, Irsoy, Lu, Dabravolski, Dredze, Gehrmann, Kambadur,
  Rosenberg, and Mann]{wu2023bloomberggpt}
Shijie Wu, Ozan Irsoy, Steven Lu, Vadim Dabravolski, Mark Dredze, Sebastian
  Gehrmann, Prabhanjan Kambadur, David Rosenberg, and Gideon Mann.
\newblock Bloomberggpt: A large language model for finance, 2023.

\bibitem[Young(1974)]{checkpoint-times}
John~W. Young.
\newblock A first order approximation to the optimum checkpoint interval.
\newblock \emph{Commun. ACM}, 17\penalty0 (9):\penalty0 530–531, sep 1974.
\newblock ISSN 0001-0782.
\newblock \doi{10.1145/361147.361115}.
\newblock URL \url{https://doi.org/10.1145/361147.361115}.

\end{thebibliography}
\bibliographystyle{colm2024_conference}
\end{document}